%% file: loop-momentum.tex
\documentclass[11pt]{article}
\usepackage{jheppub}

\def\d{\partial_z}
\def\db{\partial_{\bar z}}
\renewcommand{\Im}{\,\textrm{Im}\,}

\def\bb{\langle\!\langle\,}
\def\rr{\,\rangle\!\rangle}
\def\ab{\left[ \begin{smallmatrix} {\beta}\\
      {\alpha} \end{smallmatrix} \right]}
\preprint{CERN-TH-2018-282}

\begin{document}
\title{On integrands and loop momentum in string and field theory}
\author[]{Piotr Tourkine}
\affiliation[]{CERN, Theory Group, Esplanade des particules, 1, 1211 Gen\`eve, Switzerland}
\emailAdd{piotr.tourkine@cern.ch}

\abstract{The notion of a unique integrand does not a priori makes
  sense in field theory: different Feynman diagrams have different
  loop momenta and there should be no reason to compare them.
  In string theory, however, a global integrand is natural and allows,
  for instance, to make explicit the separation between left and
  right-moving degrees of freedom.

  However, the significance of this integrand had not really been
  investigated so far. It is even more important in view of the
  recently discovered loop monodromies that are related to the duality
  between color and kinematics in gauge and gravity loop amplitudes.
  
  This paper intends to start filling this gap, by presenting a
  careful definition of the loop momentum in string theory, and
  describing precisely the resulting global integrand obtained in the
  field theory limit.
  We will then apply this technology to write down some monodromy
  relations at two and three loops, and make contact with the
  color/kinematics duality.}

\maketitle

\section{Introduction}
\label{sec:introduction}

In the last few years, a variety of results for scattering amplitudes
in field theory at loop-level have been derived using string theoretic
methods. Interestingly, many of them have focused on integrands and
have involved explicit dependence on a loop momentum defined globally
for a string integrand.

While this is a peculiar idea from a traditional Feynman perspective,
this concept is actually present, though maybe not emphasised, since
the very early days of string theory \cite{Shapiro:1972ph}. The
seminal papers
\cite{Verlinde:1986kw,Verlinde:1987sd,DHoker:1988pdl,DHoker:1989cxq}
then laid the foundations for the definition of the loop momentum in
string theory amplitudes in their modern formulation as conformal
field theory correlation functions integrated over the moduli space of
Riemann surfaces. Those correlation functions can be written as
holomorphic squares in loop amplitudes only in presence of loop
momentum. Especially non-trivial for superstrings (in the RNS
formulation), this property was called \textit{chiral splitting}.

However, some aspects related to the precise definition of the loop
momentum had not been worked out and the recent results alluded to
above require to now re-investigate this question.
I have especially in mind two categories of results: the monodromy
relations at higher loops in string theory derived in
\cite{Tourkine:2016bak} and the scattering equation or ambitwistor
string methods at loop level.
This paper will be focussed on the former, I allude to the latter in
the discussion.

The monodromy relations in string theory were originally derived at
tree-level
\cite{Plahte:1970wy,BjerrumBohr:2009rd,Stieberger:2009hq}. They are
now understood to generalize of the Bern-Carrasco-Johanson
\cite{Bern:2008qj} (BCJ) duality between colour and kinematics that
underlie the so-called double-copy construction \cite{Bern:2010ue} of
gravity integrands as squares of Yang-Mills integrands. While
implemented very efficiently to compute loop amplitudes, see for
instance the last recent achievement at five loops and references
therein \cite{Bern:2018jmv}, this duality is still not understood from
first principles.

The tree-level relations were extended to all loop orders in
\cite{Tourkine:2016bak} in open string
theory.\footnote{This generalized some previous works in field theory
  \cite{Feng:2011fja,Feng:2010my,Boels:2011tp,Boels:2011mn,Du:2012mt}}
This gives hope that string theory can shed light on the
colour-kinematics duality and these relations need to be understood deeper.
In particular, some aspects related to the definition of the loop
momentum were only conjectured in \cite{Tourkine:2016bak} and the
present paper intends to fill this gap and show in details how to
apply the monodromy relations at higher loops.

Another aspect that this paper deals with is the notion of an
integrand in field theory. In~\cite{Tourkine:2016bak}, it was
emphasized that the relations induced in field theory by the stringy
monodromies are valid \textit{globally} at the integrand level,
i.e. mix different integrands of different graphs at the same value of
the loop momentum, as in \cite{Chiodaroli:2017ngp}.
We will see how this picture generically emerges from the field theory
limit of string amplitudes.

Here is a summary of the main contributions of this paper:
\begin{enumerate}
\item A precise definition of the loop momentum in the string theory
  integrand in \ref{sec:closed-strings}, from a review of classic
  computations and from solving directly the classical equations of
  motion for the string. The definition requires working on a
  so-called canonical dissection of the surface (see
  fig.~\ref{fig:octogon}), which, importantly, breaks modular
  invariance \cite{DHoker:1988pdl} because it does not allow to modify
  the homology basis anymore.\footnote{Only after the loop momentum is
    integrated out the invariance is restored.}
\item A careful study of its field theory limit (in
  sec.~\ref{sec:closed-strings}, which as a by-product gives how the
  loop momentum is distributed across all Feynman graphs appearing in
  this limit. This analysis uses some tools to study the degeneration
  of Riemann surfaces.
\item Finally I provide applications of these definitions in loop
  amplitudes. In particular, we shall see in details how the monodromy
  relations work two and three-loop amplitudes, which support further
  the claim that the monodromy relations generalize the BCJ
  duality. More precisely it will support the conjecture that in all
  higher loop relations, the monodromy relations always combine the
  numerators appearing in the field theory limit into groups of graphs
  called BCJ triplets.
\end{enumerate}

It should be noted that in this paper we will exclusively be concerned
with the bosonic part of the string amplitudes, which is the one that
carries the loop-momentum zero modes.

Further applications of these results are presented in the discussion
\ref{sec:discussion} together with open questions.

\section{String theory}
\label{sec:closed-strings}

The presence of loop momentum is standard in the operator formalism of
the string theory, this is for instance the was that amplitudes are
derived in the classic book by Green Schwarz and Witten
\cite{Green:2012pqa}. These representations have the advantage to make
\textit{chiral splitting} manifest \cite{DHoker:1989cxq}, i.e. the
string integrand is factorized as a product of a purely left-moving
(holomorphic) and right-moving (anti-holomorphic) part. The
traditional form of the string amplitudes is obtained after
integrating out the loop-momentum, which induces non-holomorphy in the
integrand and destroys its chiral splitting. 

The drawback, however, is that this formalism is difficult to use
at high multiplicity and loop orders because it amounts to do a very
complicated Feynman diagram computation, and the number of graphs
increases quickly. Besides, the structure of the moduli space of
Riemann surfaces at higher genus essentially renders the whole process
unusable.
The modern approach to string theory scattering amplitudes is based on
complex (super)-geometry and conformal field theory techniques
\cite{DHoker:1988pdl}.
In this manner, the non-holomorphic terms are generated from the
start~\cite{Verlinde:1986kw,Verlinde:1987sd,DHoker:1988pdl,DHoker:1989cxq},
essentially because meromorphic functions on Riemann surfaces must
have the sum of their residues vanishing (via Stoke's
theorem).\footnote{Another intuitive picture is that one cannot put a
  single electric charge at rest on a compact Riemann surface; a
  second one needs to be added to cancel the charge or a background
  charge should be included. This background charge breaks the
  holomorphy of the Green's function.} We review this construction
now.

We will review the construction of the universal part to string
theory amplitudes at loop-level. It is the generalisation of the
Koba-Nielsen factor
$\prod_{i<j} |z_i-z_j|^{\alpha' k_i\cdot k_j}$ ubiquitous to
tree-level string amplitudes. Here and throughout, $z_i$ will be the
locations of the vertex operators on the string worldsheet, $\alpha'$
is the string Regge slope, and $k_i$ are null momenta of the states,
all taken to be incoming, that satisfy momentum conservation
$\sum_{i=1}^n k_i=0$ for and $n$-particle process.

Along the way we shall see how the loop momentum appears. We will
mostly follow \cite{DHoker:1988pdl}, and supplement the construction
with careful normalisations and definitions of the loop-momentum. Note
that the paper \cite{Skliros:2016fqs} presents details on these
computations and an exhaustive reference list on the matter.

In the conformal gauge, the Polyakov action for closed strings reads
\begin{equation}
  \label{eq:Spolyakov}
  S_{\mathrm{P}} = \frac1{2\pi \alpha'}\int_\Sigma \partial_z X 
  \partial_{\bar z} X\,.
\end{equation}
where $X^\mu(z,\bar z)$ are the coordinates of the string in
$d$-dimensional target flat space.

The equations of motion of the theory without vertex operator
insertions split the $X$ field into left and right-movers as
\begin{equation}
  \label{eq:LR}
  X^\mu(z,\bar z) =X_L^\mu(z)+X_R^\mu(\bar z)
\end{equation}
which will share a common zero mode $x_L=x_R$ and a loop momentum zero
mode to be introduced momentarily. 

In the presence of $n$ standard exponential vertex operator
insertions,\footnote{Typical vertex operators would also have a
  polynomial dependence on $\partial_z X$, ghost fields, and other
  matter fields, in generic string models.}
\begin{equation}
  \label{eq:planve-wave}
  V_j(k_j) = \exp(i k_j\cdot X(z_j,\bar z_j))
\end{equation}
the phases can be inserted in the action and the object we seek to
compute is given by:
\begin{equation}
  \bb V_1(k_1)\dots V_n(k_n)\rr=\int \mathcal{D}X
  e^{
      -\frac1{2\pi\alpha'} \int d^2 z \d X^\mu \db X_\mu + 2i\pi\alpha'\sum_{j=1}^n k_j^\mu  X_\mu(z,\bar
    z) \delta^2(z-z_i) d^2z 
}
  \label{eq:path-integral}
\end{equation}
where the double bracket notation is that of \cite{DHoker:1988pdl}.
To compute this path integral, we need to invert the kinetic operator
$\d\db$, i.e. compute the Green's function
\begin{equation}
  {G}(z,w) \eta^{\mu\nu}=\langle X(z,\bar z)^\mu X(w,\bar w)^\nu\rangle \label{eq:G-def}\,.
\end{equation}
The subtlety when doing this directly comes from the fact that $\d$
and $\db$ have zero modes on a compact Riemann surface $\Sigma$ of
genus $g\geq1$, that correspond to loop momentum. They are supported
by $g$ holomorphic and anti-holomorphic one-forms $\omega_I$ and
$\bar \omega_J$ that span the cohomologies $H^{(1,0)}(\Sigma)$ and
$H^{(0,1)}(\Sigma)$ :
\begin{equation}
  \forall I=1\dots g,\quad  \partial \bar \omega_I=0,\quad
  \bar\partial \omega_I = 0\,.
  \label{eq:zm}
\end{equation}
In this equation and below, $\partial$ and $\bar\partial$ are
operators $\partial = (\partial/\partial z) dz $ and
$\bar \partial = (\partial/\partial \bar z) d\bar z$, as defined in
appendix \ref{sec:definitions}. We also abbreviate $\d :=
(\partial/\partial z)$ and likewise for $\db$.

The holomorphic one-forms are dual to a homology of one-cycles,
traditionally called $a$ and $b$ cycles, canonically defined by their
intersection numbers $a_I\cap b_J=\delta_{I,J}$, for $I,J=1\dots g$,
all other vanishing.
Pairing a cycle with a form is done via the period map $  (\omega,c) \mapsto \int_c \omega$.
Normalising the period of the 1-forms on the $a_I$ cycles to
$\delta_{IJ}$ makes the periods along the $b$ cycles define the \textit{period
matrix} $\Omega$ of the surface as follows
\begin{equation}\label{e:norm-holdiff}
 \oint_{a_I} \omega_J = \delta_{IJ}\,, \qquad \oint_{b_I}
\omega_J =
\Omega_{IJ}\,.
\end{equation}
It is a symmetric $g\times g$ matrix with positive-definite imaginary
part $\Im \Omega >0$.

Let us then fix a Riemann surface $\Sigma$ of genus $g$. The kinetic
operator can be inverted on the space orthogonal to the zero modes
\cite{Verlinde:1986kw,Verlinde:1987sd,DHoker:1988pdl} and the
equations that define the corresponding Green's function are
\begin{align}
  &\int_\Sigma G(z,w)d^2z  =0\,,\\
  &\d \db G(z,w) = -2\pi \alpha' \delta^{2}(z-w)+\frac{2\pi\alpha'}{\int
    d^2z \sqrt{g}}\,,\\
  &\d \partial_{\bar w} = 2\pi\alpha' \delta^{(2)}(z-w) -{
    \alpha'\pi}\sum _{I,J} \omega_I(z)(\Im \Omega)^{-1}_{IJ}\bar \omega_J(w)\,.
\end{align}
where $g$ is the determinant of the metric on the surface, as defined
in \ref{sec:definitions}.
These equations can be solved and yield
\begin{equation}
{G}(z_1,z_2)= -\frac {\alpha'}2 \ln\left(|E(z_1,z_2)|^2\right) +\alpha'\pi
\Im \left(\int_{z_2}^{z_1} {
\omega_I}\right) ((\Im \Omega) ^{-1})^{IJ} \Im\left(\int_{z_2}^{z_1} { \omega_J}
\right)\,.
\label{eq:bosprop}
\end{equation}
up to terms which we neglect because vanish on the support of momentum
conservation.
The \textit{prime form} $E$ is defined in (\ref{e:primeform}). Its
essential property is that it vanishes linearly on the diagonal
$$E(x,y)=x-y+O(x-y)^3.$$
It is defined on the universal cover of $\Sigma$, because it has
monodromies (given in eq.~(\ref{e:multE})) along $a$ and $b$ cycles
transportation. The non-holomorphic correction in
eq.~(\ref{eq:bosprop}) exactly cancels these monodromies and the
Green's function is correctly defined on the surface and not its
cover.

The correlation function (\ref{eq:path-integral}) is then computed by
Wick's theorem:
\begin{equation}
\langle \prod_{i=1}^{n}e^{i k_i X(z_i,\bar  z_i)}  \rangle = e^{
  - \sum_{i<j} k_i \cdot k_j G(z_i,z_j) }\label{eq:wick}
\end{equation}
Because of the non-holomorphic terms, this expression cannot be
written as it stands as a modulus square.
Note that they are absent at tree-level,
\begin{equation}
\langle X^\mu(z_1,\bar z_1) X^\nu(z_2,\bar z_2)\rangle  =
-\frac{\alpha'}{2}\eta^{\mu\nu} \ln(|z_1-z_2|^2)\label{eq:G-tree}
\end{equation}
and the correlator \eqref{eq:wick} can be chirally split. At loops,
where the $\ln|E|^2$ term similarly poses no problem: in the
exponential of (\ref{eq:wick}) the problematic terms are
\begin{equation}
  \label{eq:zm-KN}
 Q_{\mathrm{NH}}=
  \alpha'\pi \sum_{i<j} k_i \cdot k_j \Im \left(\int_{z_j}^{z_i} {
\omega_I}\right) ((\Im \Omega) ^{-1})^{IJ}
\Im\left(\int_{z_j}^{z_i} { \omega_J}
\right)
\end{equation}

\begin{figure}[t]
  \centering
  \def\svgwidth{350pt}
  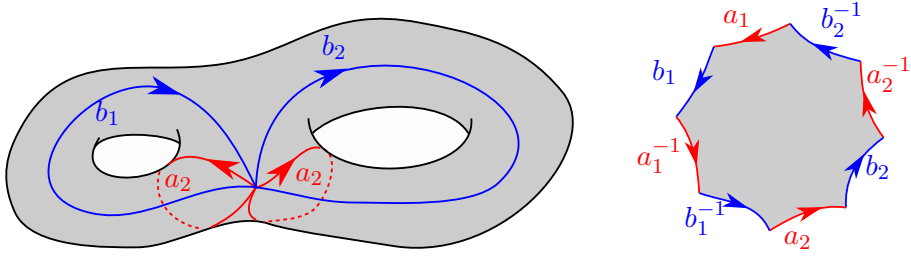
  \caption{Canonical dissection along the homology cycles.
  }
  \label{fig:octogon}
\end{figure}

Let $P$ be a point on the surface, so that we can decompose the
integration $\int_{z_j}^{z_i}$ as $\int_{z_j}^{P}+\int_{P}^{z_i}$,
\eqref{eq:zm-KN} then becomes
\begin{equation}
  \label{eq:zm-KM-P}
  Q_{NH}=\alpha'\pi\sum_{i<j} k_i \cdot k_j
  \Im \left(\int_{z_j}^{P} + \int_P^{z_i} {\omega_I}\right)
  ((\Im \Omega) ^{-1})^{IJ}
  \Im\left(\int_{z_j}^{P}+\int _P^{z_i} { \omega_J}\right)
\end{equation}
The diagonal terms
$\Im \int_P^{z_i}\omega_I ((\Im \Omega)^{-1})^{IJ} \Im \int_P^{z_i}
\omega_J$ vanish by momentum conservation (summing over $j$ in this
case), so we keep only the crossed terms and we would want to rewrite~\eqref{eq:zm-KM-P} as
\begin{equation}
  Q_{NH}= -2 \pi \alpha'\sum_{i<j} k_i \cdot k_j \Im \left(\int_{P}^{z_j}  {\omega_I}\right)
  ((\Im \Omega )^{-1})^{IJ}
  \Im\left(\int _P^{z_i} { \omega_J}\right)
  \label{eq:equad-form}
\end{equation}
(the sign comes from flipping the orientation of the integration in
one term). The reason why this identity is not straightforward is
because it is valid if and only if all the paths from $P$ to $z_i$
need to be uniquely defined. Hence, we are looking for a way to define
uniquely, for all values of $z_i$ on $\Sigma$, a path from $P$ to
$z_i$. Ambiguities can arise from $z$ winding along a non-trivial
cycles, and therefore what we describe is a way to cut open the
Riemann surface into a polygon with $4g$ faces, called its canonical
dissection, as in fig.~\ref{fig:octogon}.
It is defined by cutting open the surface along the $a$ and $b$
cycles, not considered anymore as representatives in the homology, but
as actual curves, all of which touching in one point exactly.

Now, because the sum is factorized in \eqref{eq:equad-form},
we can introduce a Gaussian $d$-dimensional integration so that
\begin{multline}
  \label{eq:gaussian}
 (\det \Im \Omega)^{d/2} \int \frac{d^D \ell }{(2\pi)^D} e^{ -
   \pi\alpha' \ell _I \ell _J \Im \Omega^{IJ}  - 2\pi\alpha' \sum_{i,J} \ell_J  \cdot k_i \Im \int_P^{z_i} \omega_J}
=\\
e^{2\pi \alpha'\sum_{i<j} k_i \cdot k_j \Im (\int^{z_j}_{P}  {\omega_I})  ((\Im \Omega )^{-1})^{IJ}  \Im(\int _P^{z_i} { \omega_J})}
\end{multline}
up to a global normalisation factor.
Using that $\exp(-2\Im x)=|\exp(x)|^2 $, this expression can be
further rewritten as a modulus square, and finally we have
\begin{equation}
\bb V_1(k_1)\dots V_n(k_n)\rr = \int \frac{d^d\ell}{(2\pi)^d} \bb V_1(k_1)\dots V_n(k_n)\rr
(\ell_I)
\end{equation}
where
\begin{equation}
 \bb V_1(k_1)\dots V_n(k_n)(\ell_I)\rr
  =
  (\Im \Omega)^{d/2}\bigg|
   e^{ \tfrac{i \pi \alpha'}{2}\ell _I \ell _J  \Omega^{IJ}  + i\pi\alpha'\sum_{i,J} \ell_J
     \cdot k_i  \int_P^{z_i} \omega_J}
   \prod_{i,j}{E(z_i,z_j)}^{\alpha' k_i k_j/2}\bigg|^2\,\,
   \label{eq:corr-fun}
\end{equation}
which is eq.(2.99) of \cite{DHoker:1988pdl}. This is the content of
chiral splitting for the bosonic part of the amplitudes.

The most important conclusion of this section is that the loop momenta
are defined with respect to a specific canonical dissection, and not
just the homology. Now we will see how this can be derived from
looking at the classical trajectory for the field $X$; this will lead
to a precise definition of the momentum flowing through a given cycle.

A consequence of working on a canonical dissection is that modular
invariance (the freedom to change $a$ and $b$ cycles) is totally
broken, because $a$ and $b$ cycles cannot be mixed anymore within the
string integrand. Of course, re-integrating out the loop momentum
gives modular invariant expressions.

\paragraph{Classical solution.}
Since the action is free, all those quantities could have been
equivalently computed from the classical solution of the
Euler-Lagrange equation with sources. For $X$, it can be obtained by
varying the action \eqref{eq:path-integral} or, equivalently, by
computing
\begin{equation}
  \label{eq:X-class}
  X_{\mathrm{class}}^\mu(z,\bar z)=\frac{\langle X^\mu(z,\bar z) \prod_{i=1}^{n}e^{i k_i
    X(z_i,\bar  z_i)}
  \rangle}{\langle \prod_{i=1}^{n}e^{i k_i
    X(z_i,\bar  z_i)}
  \rangle}
\end{equation}
using the individual two-point functions $\langle X X\rangle$.

Let us follow the former approach. We want to minimize the following action
\begin{equation}
  \label{eq:S_exp}
  S=\frac{1}{2\pi\alpha'}\int_\Sigma \partial_z X 
  \partial_{\bar z} X + 2i\pi \alpha' \sum_j k_j^\mu X(z,\bar z)
  \delta^2(z-z_i,\bar z-\bar z_i) \,.
\end{equation}

It is instructive to first do the computation at tree-level
when there are not yet zero modes.
The $\frac{\delta}{\delta X}$ variation of this Lagrangian yields
\begin{equation}
  \label{eq:classical}
  2\d
  \db X^\mu
  =2i \pi \alpha'\sum_i k_i^\mu\delta^2(z-z_i,\bar
  z-\bar z_i) 
\end{equation}
Using that
\begin{equation}
\db \frac{1}{z}= \d \frac 1{\bar z}=2\pi \delta^{(2)}(z,\bar
z)\,,
\label{eq:delta-def}
\end{equation}
this integrates once to
\begin{equation}
  \db X^\mu
  =\frac{i\alpha'}{2}\sum_i \frac{k_i^\mu }{\bar z-\bar z_i}
\end{equation}
and then
\begin{equation}
  X_\mathrm{class}^\mu
  =x_R +\frac{i\alpha'}{2}\sum_i k_i^\mu \ln(\bar z-\bar z_i) + X_L(z)\,.
\end{equation}
The holomorphic part is determined by re-injecting this equation in the
equations of motion and one finds
\begin{equation}
  \label{eq:Xclasstree}
  X_\mathrm{class}^{\mu} = x_0+ \frac{i\alpha'}{2}
  \sum_i \ln|z-z_i|^2
\end{equation}
where $x_0=x_L + x_R$ is the zero-mode that gives rise to momentum
conservation upon $\int d^d x_0$ integration.

Let us now go to loop level and consider a Riemann surface $\Sigma$ of
genus $g$.
Analogously to the tree-level case, we can obtain the singular part of
$ \partial X^\mu_L(z)$ in terms of meromorphic differentials with
single poles $\omega_{z_+,z_-}=\omega_{z_+,z_-}(z)dz$ with residue $\pm1$ at
$z=z_\pm$. They are called \textit{abelian differentials of third
  kind}\footnote{For a standard reference, see \cite{farkas}.} and can
be uniquely defined by normalising to zero their periods along the
$a$-cycles:
\begin{equation}
  \label{eq:third-kind}
  \omega_{z_+,z_-}(z)\underset{z\to
    z_\pm}{\sim}\pm \frac{1}{z-z_\pm};\qquad \forall I=1\dots g,\quad \oint_{a_I} \omega_{z_+,z_-} = 0\,.
\end{equation}
For further convenience, let us denote $c_i$ the circles
$|z-z_i|=\epsilon$. This allows to define a singular homology on
$\Sigma-\{z_1,\dots,z_n\}$ by augmenting that of $\Sigma$ with these
$n$ new $c$ cycles.

The new ingredient compared to the tree-level case is the presence of
zero-modes for the $\bar \partial$ and $ \partial $ operators, given by
the holomorphic one-forms and their complex conjugates, as in~(\ref{eq:zm}).
After the first integration of \eqref{eq:classical}, we find
\begin{align}
  \partial X^\mu_L(z) = i\pi\alpha' \sum_{J=1}^{g}\omega_J \ell _J^\mu +
  \frac{i\alpha'}{2}\sum_{i=1}^{n}\omega_{z_i,z_0} k_i^\mu
  \label{eq:dX}\\
\bar  \partial X^\mu_R(\bar z) = i\pi\alpha' \sum_{J=1}^{g}\bar
  \omega_J\tilde \ell _J^\mu
  +  \frac{i\alpha'}{2}
  \sum_{i=1}^{n}\bar \omega_{z_i,z_0}k_i^\mu
  \label{eq:dXbar}
\end{align}
where $z_0$ is an extra variable whose dependence drops out by
momentum conservation. I left unspecified the zero modes for the
holomorphic and anti-holomorphic fields, they will be fixed later by
physical requirement of measure a correctly normalised momentum.
Integrating once more gives
\begin{align}
  \label{eq:XclassL}
  X_{L,\mathrm{class}}^\mu(z)=  x_L^\mu+i\pi \alpha' \sum_{J=1}^{g} \ell _J^\mu \int_P^z \omega_J +
   \frac{i\alpha'}{2} \sum_{i=1}^{n}k_i^\mu\int_P^z\omega_{z_i,z_0} \\
  \label{eq:XclassR}
  X_{R,\mathrm{class}}^\mu(\bar z)=  x_R^\mu+
  i\pi  \alpha'\sum_{J=1}^{g} \tilde \ell _J^\mu
  \int_P^{\bar z}\bar  \omega_J +
  \frac{i\alpha'}{2}\sum_{i=1}^{n}k_i^\mu\int_P^{\bar z}\bar \omega_{z_i,z_0} 
\end{align}
Finally, $X_\mathrm{class}^\mu$ is given by the sum of these two
equations. To make contact with the previous derivation and
eq.~(\ref{eq:corr-fun}) in particular, note that the prime form is
related to the abelian differentials of the third kind by
\begin{equation}
  \label{eq:third-prime}
  \d \ln\left(\frac{E(z,a)}{E(z,b)}\right) = \omega_{a,b}(z)
\end{equation}

This also defines uniquely the zero modes of
$\partial X, \bar \partial X $ with correct normalisation. To measure
the loop momentum flowing through a typical cycle $C$, which is a
combination of the canonical $a_J$ cycles and $c_i$ cycles, we define
the following flux
  \begin{equation}
  \label{eq:zero-mode}
  P_C^\mu = \frac{1}{2\pi\alpha'}\oint_{C} (-\partial_zdz +\db d\bar z) X
\end{equation}

The normalisation is fixed in a first stage by demanding that
integration along $c_i$ cycles provides momentum $k_i$:
\begin{equation}
  \label{eq:zero-mode-k}
  \frac{1}{2\pi\alpha'}\oint_{c_i} (-\partial_zdz +\db d\bar z) X^\mu
  =
  k_i^\mu\frac{i\alpha'}{4\pi\alpha'}\oint_{c_i}(-\omega_{z_i,z_0}+\bar
  \omega_{z_i,z_0}) = k_i^\mu
\end{equation}
Then we have
\begin{equation}
  \label{eq:zero-mode-a}
  \frac{1}{2\pi\alpha'}\oint_{a_I} (-dz\partial_z + d\bar z \db) X^\mu =
  -\delta_{IJ} i(\ell_J^\mu-\tilde \ell_J^\mu)/2 \equiv \ell_I^\mu
\end{equation}
if the loop momenta are taken to be purely imaginary. This derivation
gives another check of this property which was originally observed in
\cite{DHoker:1988pdl,DHoker:1989cxq} and that seems fundamental to
string theory on euclidean worldsheets. It would be interesting to
study the consequences of this fact in the ambitwistor string where a
similar normalisation was observed to arise by matching against field
theory computations in \cite{Casali:2014hfa}.

Open strings on orientable surfaces are obtained by modding out by the
involution $z\simeq \bar z$ along the $a$-cycles of the string
worldsheet~\cite{Bianchi:1989du} and letting the punctures live on the
boundary of the surfaces. More precisely, if $z=\rho e^{i\theta}$ with
$\theta\in[0,2\pi[$ is a local coordinate along an $a$ cycle, we
identify $\theta\leftrightarrow -\theta$. This is the natural
involution to describe the gauge theory channel of open string
amplitudes, which we will use later to apply the monodromy relations
in open string theory and their induced relations in field
theory. This involution can also be used to obtain some non-planar
graphs, as long as they are given by orientable surfaces. 

Note also that this turns the cycles of the canonical dissection into
segments on the worldsheet such that $\oint_{a_I} \omega_J
\to\int_{a_I'} \omega_J = \delta_{IJ}/2$ where $a_I'$  is $a_I$ modulo
the involution.

\section{Field theory integrand.}
\label{sec:field-theory-limit}

In this section we will investigate one implication of the previous
considerations. Since there exists a global integrand in string
theory, there needs to exist one in field theory, induced via the
field theory limit.
In practice, after studying the field theory limit itself, we will be
able to describe the graph integrand topologies: external leg ordering,
and labeling of the internal loop momenta.

The understanding of the mechanism of the field theory limit of
string graphs is almost as old as string theory itself
\cite{Scherk:1971xy}. 
It is produced by corners of the moduli space where the surface
degenerate so that all internal edges become infinitely long and thin
(this is a $b$-cycle statement) or equivalently where all $a$ and
$c$-cycles are pinched. This is a \textit{continuous} process, known
in the maths literature as a tropical limit \cite{Tourkine:2013rda}.

The property which we will need to describe the graphs
loop-momentum-labeling is that the momentum flowing through a cycle is
preserved by the field theory limit. As the momentum is a zero-mode,
it is not affected by the decoupling of the excited states of the
string, therefore the result which we seek for is physically sound and
the problem reduces to a computational matter.
\begin{figure}[t]
  \centering
    \def\svgwidth{250pt}
    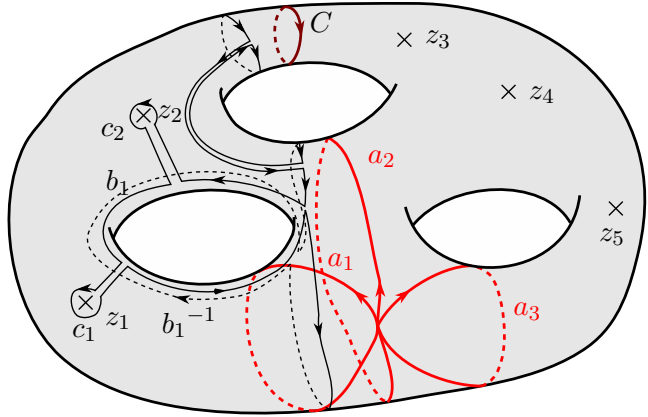
  \caption{From the picture we see that $C \cup (a_2)^{-1} \cup b_1
    \cup (b_1)^{-1} \cup c_1 \cup c_2 =  \mathrm{id}$, hence $C=a_2
    \cup (c_1)^{-1}\cup (c_2)^{-1}$ and the momentum flowing through
    $C$ is given by $\frac1{2\pi\alpha'}\oint_C (-\partial +\bar \partial)X^\mu = \ell_2^\mu-k_1^\mu-k_2^\mu$.}
  \label{fig:3-loop-example}
\end{figure}
Let $C$ be a closed curve made of $a_I$- and
$c_i$-cycles:
\begin{equation}
  \label{eq:C-def}
  C=\cup_{i\in \mathcal{I}_C} \mathbf{a}_i
\end{equation}
where $\mathbf{a}$ is either a $a$ cycle or a $c$ cycle with
coefficient $1$. This defines implicitly the set $\mathcal{I}_C$. This
excludes the possibility that our cycle $C$ could wind multiple
times. For illustrative purposes, see
fig.~\ref{fig:3-loop-example}. 
Let us call the corresponding momentum
\begin{equation}
  \label{eq:pC}
  p_C^\mu = -\frac{1}{2\pi\alpha'}\oint_C (dz \d -d\bar z \db
  )X=\sum_{I,i\in \mathcal{I}_C}
  (\ell_I^\mu + k_i^\mu)
\end{equation}
with obvious notations for the summation. The crucial point is that
this quantity is a topological invariant, therefore it cannot change
as we deform continuously the surface when taking the field theory
limit.
We now will check this property and see that when the $C$ cycle
degenerates, as in fig.~\ref{fig:loops} and show that a propagator
$1/p_C^2$ factorizes out of the string amplitude.

\subsection{Single pinching of a Riemann surface.}
\label{sec:field-theory-limit-starter}
\begin{figure}[t]
  \centering
  \def\svgwidth{450pt}
  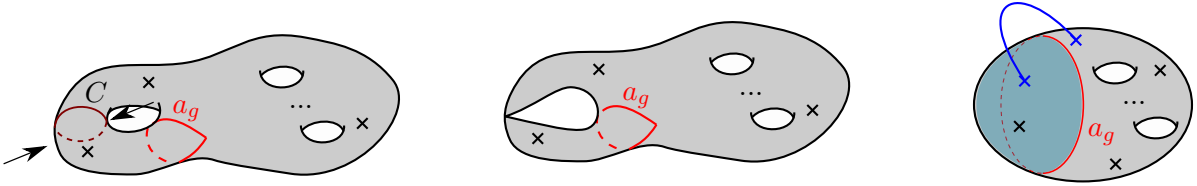
  \label{fig:loops}
  \caption{Handle-representation drawing of the pinching that we
    studied in this section.}
\end{figure}

There are two types of degenerations that a Riemann surface can
undergo: separating and non-separating. The separating degeneration
corresponds to pinching off a trivial cycle in the homology or a
$c$-cycle: it splits appart a surface of genus $g$ into two surfaces
of genera $g_1$ and $g_2$ such that $g=g_1+g_2$. A non-separating
degeneration pinches off an $a$-type cycle and the resulting object is
a surface with genus decreased by one unit and two extra
punctures. This is the case of interest for us because we want to
check that a propagator with expected loop momentum labeling is
generated. An example of such a degeneration is provided in fig.\ref{fig:loops}.

Firstly, let us observe that we do not loose in generality by
considering that the $a$ cycle part of our $C$ cycle is the cycle
$a_g$ (we could always relabel the $a$ cycles). 

The degeneration of the Riemann surface is done via the so-called
\textit{plumbing fixture} construction see \cite{Fay} or
\cite{DHoker:1988pdl,DHoker:2017pvk}.
In this construction, the degenerating curve $\Sigma_g$ is constructed
from a Riemann surface of genus $g-1$, $\Sigma_{g-1}$ with period
matrix $\Omega_{g-1}$ and a pair of points marked on the surface $p_a$
and $p_b$. To construct $\Sigma_g$, one constructs two pairs of
circles centered around $p_a$ and $p_b$: $C_a''$ and $C_b$' of radius
1 and $C_a',C_b''$ of radius $|q|<1$ for complex number $q$ that
parametrizes the degeneration. The internal disk is then cut out of
the surface and the annuli between the disks are identified via an
invertible map $$xy = q\,.$$
The extra $a$ cycle $a_g$ is a closed loop around $C_a''$ for
instance; the extra $b$ cycle $b_g$ is a line that connects any two points
$z_a$ and $z_b$ in the annuli that obey $z_a z_b=q$. Choosing
$|z_a|=|z_b|=\sqrt{|q|}$ ensures that when $q=0$, the extra cycle is
  really the line connecting the two points $p_a$ and$p_b$ which are identified.
\begin{figure}[t]
  \centering
  \def\svgwidth{300pt}
  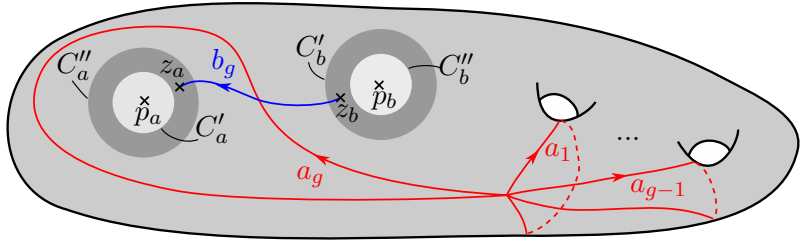
  \caption{Illustration of the plumbing fixture construction.}
  \label{fig:plumbing}
\end{figure}
Then, if $\Omega_g$ is the period matrix of $\Sigma_g$, Fay in
\cite{Fay} proves that
\begin{equation}
  \Omega_{g}\sim
\left(\begin{array}{@{}c|c@{}}
\mbox{  $\Omega_{g-1}$ \phantom{\bigg|}}
  &  \vec v \\
\hline
  \vec v {\,}^t &
\tau
      \end{array}\right)\,,\quad \mathrm{where~}q=e^{2i\pi\tau}\,,
    \label{eq:PM-degen} 
\end{equation}
up to sub-leading terms and where the components of $\vec v$ are given by
$v_I = \int_{p_a}^{p_b}\omega_I^{(g-1)}$, $I=1\dots g-1$. The exponent
on the differential form $\omega_I^{(h)}$ designate the surface
$\Sigma_h$ to which it is
associated for $h=g,g-1$.

With this, we can already extract the degeneration of the quadratic
term in the loop momentum in the exponential in
eq.\eqref{eq:corr-fun}:
\begin{equation}
  \label{eq:loop-quad-degen}
  \sum_{I,J=1}^{g}\ell_I \cdot \ell_J \Im \Omega_g = \ell_g^2 \Im\tau + 2
  \sum_{I=1}^{g-1} \ell_I \cdot\ell_g \Im(v_I)+ \sum_{I,J=1}^{g-1}\ell_I\cdot \ell_J \Im \Omega_{g-1} 
\end{equation}

To study the degeneration of the other two term in
\eqref{eq:corr-fun}, we need the degeneration of the differential
forms one-forms. That of the holomorphic forms is standard and
detailed in the references mentioned above:
\begin{align}
  &\omega_I^{(g)}=\omega_I^{(g-1)} +O(q)\\
  &\omega_g^{(g)} = \omega_{p_a,p_b}^{(g-1)}+O(q)\,.
\end{align}
This allows to extract the degeneration of the second term in
\eqref{eq:corr-fun}:
\begin{equation}
  \label{eq:degen-2nd}
\sum_{i,J} \ell_J  \cdot k_i  \int_P^{z_i}
  \omega_J^{(g)}=
  \ell_g \cdot\sum_{i=1}^n k_i  \int_P^{z_i}  \omega_{p_a,p_b}^{(g-1)} +
  \sum_{i=1}^n\sum_{J=1}^{g-1} \ell_J  \cdot k_i  \int_P^{z_i} \omega_J^{(g-1)}+O(q)
\end{equation}

\begin{figure}[t]
  \centering
  \def\svgwidth{300pt}
  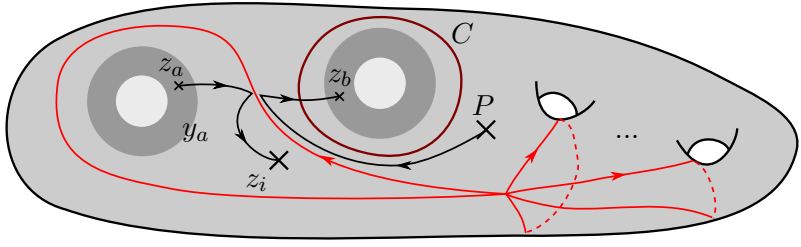
  \caption{Illustration of the plumbing fixture construction.}
  \label{fig:plumbing-pinch}
\end{figure}

We therefore need to evaluate the integrals $ \int_P^{z_i}
\omega_{p_a,p_b}^{(g-1)}$. The circle cycle $C$ defined as above is
represented on the previous picture in
figure~\ref{fig:plumbing-pinch}. It cuts out the surface into two
distinct components (we are still working in the canonical dissection
hence one should not cross through the $a$ cycles).

When $P$ and $z_i$ are on the same side, almost nothing is to be done
and $ \int_P^{z_i} \omega_{p_a,p_b}^{(g-1)}$ provides directly two
terms similar to the $k_i\cdot k_j$ terms of
eq.\eqref{eq:corr-fun}. To see this, we use the reciprocity theorem\footnote{See e.g. \cite[III.7]{farkas} -- our $\omega_{PQ}$ forms are denoted
$\tau_{PQ}$ there.}
for abelian differentials of the third kind with zero $a$ periods:
\begin{equation}
  \label{eq:reciprocity}
  \int_B^A \omega_{C,D}= \int_D^C \omega_{A,B}
\end{equation}
Therefore we have that
\begin{equation}
  \label{eq:recip-degen}
\int_P^{z_i}\omega_{p_a,p_b}^{(g-1)} =
\left(  \int_{z_0}^{p_a}-\int_{z_0}^{p_b}\right)\omega_{z_i,P}^{(g-1)} 
\end{equation}
which now has the desired form, if $z_0$ is chosen as in
\eqref{eq:XclassL}, \eqref{eq:XclassR}. While these terms have been
easy to obtain, the ones that descend from degenerating the other
terms in the exponent of (\ref{eq:corr-fun}) that contribute to induce
a new Koba-Nielsen factor on the resulting pinched-and-dissected
surface are more subtle and we shall not treat them here, but instead
focus exclusively on how the propagator $1/p_C^2$ is produced.

If now $z_i$ is on the other side of the cycle $C$, the path between
$P$ and $z_i$ is a sum of two segments, as in
fig.~\ref{fig:plumbing-pinch}:
\begin{equation}
  \int_P^{z_i}= \int_P^{z_b}+ \int_{z_a}^{z_i}
\end{equation}
As seen in the picture, the path can be deformed so as to make
apparent that it contains the following integral:
\begin{equation}
  \label{eq:b-emerge}
  \int_{z_b}^{z_a} \omega_{p_a,p_b}^{(g-1)}
\end{equation}

Generically, this term is equal to $\tau$, up to sub-leading
corrections or order $O(1)+O(q)$. If we follow the refinement of the
plumbing fixture construction developed in \cite{DHoker:2017pvk}
called the ``funnel formalism'', this integral is exactly equal to
lower right entry of the period matrix $\tau$ in
eq.~(\ref{eq:PM-degen}), see
\cite[(3.27)]{DHoker:2017pvk}.\footnote{The extra bits of the contour
  add up to create what becomes the Koba-Nielsen factor on the cut
  surface, which are not in the scope of this paper as we said above.}

If $I=\{i_1\dots i_k\}$ denotes the set of particles being on the
other side of the cycle $C$, from those terms we therefore get a
global factor of
\begin{equation}
  \label{eq:middle}
  4i\pi \tau \ell_g\cdot \sum_{i\in I} k_i
\end{equation}

Finally we need to investigate the last category of terms, those of
the form $k_i\cdot k_j \int_P^{z_j}\omega_{z_i,z_0}$. But the
degeneration is essentially identical to what we did before. When
$z_j$ is on the same side of the cycle as $P$, nothing happens. If
$z_j$ is on the other side, we get a factor of $\tau$ for each of
these $z_j$.

Equivalently, because of the equation~(\ref{eq:third-prime}) we need
the degeneration of the prime form.
In \cite{Tourkine:2013rda}, the full degeneration of the string
worldsheet integrals into worldline graphs (``tropical graphs'') was
studied and it was verified that the logarithm of the prime form
descends to the worldline propagator of \cite{Dai:2006vj}. The latter
is given by the sum of the distance in the graph between two points,
which essentially parametrizes the degeneration. In field theory, for
a graph with an edge of proper time $T$, there always is a modulus
of the Riemann surface parametrized by
$q=\exp(-2\pi(T/\alpha' + \theta)\to 0$ for
$\theta\in[0;2\pi[$ such that
\begin{equation}
  \ln(E(x,y)) = \ln(q)+...
\end{equation}
If we now look at our case where the surface is degenerated in one
cycle, this fact needs to remain true (essentially because the limit
is continuous and the deformation of this cylinder does not influence
the other moduli of the surface to first approximation) and all the
propagators $\ln(|E(z_i,z_j)|)$ that end up splitting appart two
punctures on each side of $C$ produce a factor of
$\ln(q)$.\footnote{It would be interesting to use the funnel formalism
  developed in \cite{DHoker:2017pvk} to prove this fact directly.} If
we call $\mathcal{I}$ and $\mathcal{J}$ the (disjoint) sets of
punctures on each side of the cut, we get a total factor of
\begin{equation}
  \label{eq:E-degen}
  \ln(q)\sum_{i\in \mathcal{I},j\in \mathcal{J}}(k_i\cdot k_j)
\end{equation}

To conclude, we can collect all the terms that undergo a degeneration
in \eqref{eq:corr-fun}. They conspire to produce a quadratic propagator
given by $K^2 = (\ell+\sum k_i)^2$ which appear as follows
\begin{equation}
  \label{eq:loop-mom}
  \int_{|q|<\epsilon} \frac{d^2 q}{|q|^2}|q|^{-\alpha' \pi
    K^2}\propto\frac1{\alpha' K^2}+O(\epsilon)
\end{equation}
where we have used that a $d^2\tau \propto d^2q/|q|^2$ is a modulus of
the surface, and hence is being integrated over in the full string
amplitude.  It can also be checked that all other dependence on the
modulus drops, to sub-leading order, as far as the exponential is
concerned.\footnote{Other terms may appear in front of the
  plane-waves, when scattering gravitons for instance, but their
  presence only affects the numerators of the field theory graphs, not
  the propagators.}

Using this property in combination with the observation that the cycle
running through a node is a topological invariant proves that all the
graphs obtained in the tropical or field-theory limit can be given a
uniform loop momentum. In the next section we study this labeling and
use it in the monodromy relations.

\subsection{Graph labeling in the field theory limit}
\label{sec:graph-labeling-field}

\paragraph{Closed string.}

Let us now study the graph labelling induced in the field theory limit
for this closed string picture.
The choice of the point where the cycles touch in
fig.~\ref{fig:octogon} defines all the possible degeneration channels
and the associated momenta. The graphs are obtained by letting the
puncture travel through the whole surface, and pinching all possible
$a$-type cycles. To identify the momentum flowing through an inner
edge, work out which homology cycle being pinched (as in
fig.~\ref{fig:3-loop}) on the Riemann surface and derive the momentum
flowing through it with the rules of eqs.~(\ref{eq:zero-mode-k}),
(\ref{eq:zero-mode}).

The fact that the punctures move over the whole surface implies in
particular that, for a given type of degeneration with prescribed
momenta,\textit{ all the graphs with legs permuted should appear in the
  integrand}.

\paragraph{Open string.}

In the open string, the graph labeling that emerges from the
integration over the string moduli space is similar to the closed
string picture, expect that individual graphs are color ordered. This
allows to select restricted classes of numerators when studying
properties like the monodromy relations.

The open string will be the subject of the next section where we study
the monodromy relations at two and three loops. I give there more
details and examples on the systematics of the limit and the labeling.

\begin{figure}[b]
  \centering
  \includegraphics[scale=0.4]{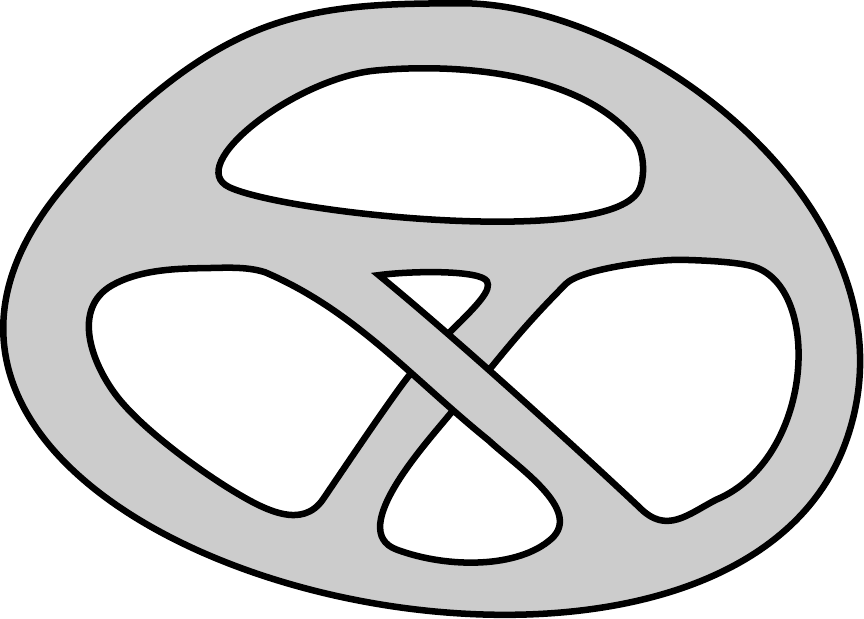}
  \caption{Non-planar-looking closed string graph.}
  \label{fig:non-planar}
\end{figure}

\paragraph{Non-planar graphs}
There are two types of non-planar contributions present in the field
theory limit of closed string graphs (in gravity amplitudes);
non-planar vacuum graphs and planar vacuum graphs where external legs
are inside. While the latter may seem to cause no troubles concerning
the definition of the loop momentum, the former may appear
problematic.
They are actually not and are neatly generated by the mechanism of the
field theory limit (pinching $a$-cycles), therefore they also come
with a uniquely defined loop-momentum. The interested readers can look
at the graph in fig.~\ref{fig:non-planar} and convince themselves that
the graph suggested by the drawing of this Riemann surface can be
obtained from a regular ``planar-looking'' genus 4 surface by pinching
a sum of $a$ cycles with $\pm1$ coefficients.

There are non-orientable open string graphs, and it would be
interesting to study the loop momentum of these graphs too.
\footnote{For a study of the monodromy relation for non-orientable at
  one loop, see \cite{Hohenegger:2017kqy}.}

\section{Application to the monodromy relations in open string and
  gauge theory at loop-level.}
\label{sec:gauge-theory-at}
Monodromy relations to all loop orders were derived in
\cite{Tourkine:2016bak} in a representation involving loop
momentum. Compared to the tree-level case
\cite{BjerrumBohr:2009rd,Stieberger:2009hq}, the relations do not hold
at the level of the amplitude but at the level of the integrand. This
stems from the fact that the integrand has both local and global
monodromies, and the latter involve phases that depend on the loop
momentum. The whole construction is fairly simple and exposed in
\cite{Tourkine:2016bak} so it will not be reviewed too deeply here.

\begin{figure}[t]
  \centering
  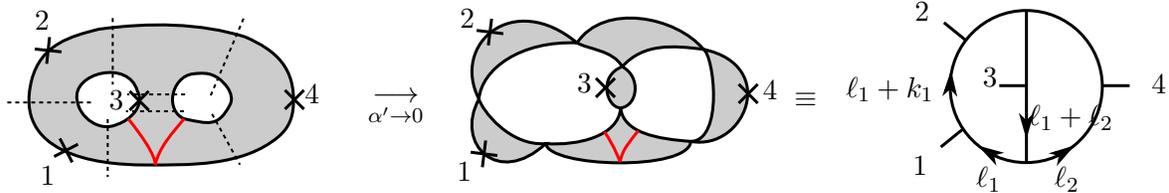
  \caption{Two-loop example of field theory limit for open strings and
    determination of the loop momentum}
  \label{fig:two-loop-ex}
\end{figure}

The basic idea is to consider a particular open-string loop-diagram
with particles ordered along the boundaries (inner and outer). Using a
representation with loop momentum yields directly an integrand that is
holomorphic, as we saw above. Therefore, taking one of these particles
along a closed contour inside the surface gives, via the residue
theorem, that the sum of all individual portion vanishes exactly at
the integrand level. Each portion can be rewritten as a properly
ordered open string integrand but at the cost of picking up a phase,
that depends on the loop momentum when the particle is on a different
boundary than the one we started from. The portions of the contour
that run along the $a$-cycles (in red in fig.~\ref{fig:two-loop-ex})
cancel after loop momentum integration (they are related by a simple
shift in the loop momentum see \cite{Ochirov:2017jby} for detailed
examples at one-loop).

\subsection{Two loops}
\label{sec:two-loops}
To be concrete, I provide an example of the field theory limit of
two-loop four-gluon amplitude in type I superstring.s

The orientable topologies of the open-string amplitude (no cross-caps)
for $\mathcal{N}=4$ Yang-Mills at two loops are obtained from the
celebrated two-loop formulae for closed strings of D'Hoker \& Phong
\cite{D'Hoker:2002gw}.
They read 
\begin{equation}
  A^{(2)}_{\mathrm{orient}} (1,2,3,4)
  =
  s_{12}s_{23} A^{\mathrm{tree}}(1,2,3,4)
  \int \frac{\prod_{I\leq J} d \Omega_{IJ}}
  {(\det\,{\Im}\,{\Omega})^5}
  \int_{(\partial\Sigma)^4}
  {\mathcal Y}_S
\exp(\sum_{i,j}
k_i \cdot k_j \mathcal{G}(z_{ij}))\,.
\label{e:4gravgen2}
\end{equation}
up to a global normalisation factor, and where
$A^{\mathrm{tree}}(1,2,3,4)$ is the tree-level four-gluon
colour-ordered partial amplitude, while the kinematics invariants are
defined by $s_{ij}=-(k_i+k_j)^2$. The integration ordered along the
boundary $(\partial\Sigma)^4\simeq \{\forall i=1\dots 4, z_i\in \partial \Sigma , z_1<z_2<z_3<z_4\}$ and
$\mathcal{Y}_S$ is defined by
\begin{equation}\label{e:Ys}
3  \mathcal Y_S= (k_1-k_2)\cdot (k_3-k_4)\, \Delta(z_1,z_2)\Delta(z_3,z_4)+
(13)(24) + (14)(23)\,,
\end{equation}
in terms of the differential forms bilinears 
\begin{equation}
  \Delta(z,w)=\omega_1  (z) \omega_2(w)-\omega_1(w) \omega_2(z)\,,
\label{e:Delta}
\end{equation}

For maximally supersymmetric amplitudes in general at two loops in
type I and type II, the field theory limit procedure was worked out
in~\cite{Green:2008bf,Tourkine:2013rda} and the numerators are given
by the tropical limit of the factor $\mathcal{Y}_S$, which equals the
kinematic invariant $s_{ij}$ whenever the two legs are on the same
$b$-cycle and no sub-triangle is present in the graph, which matches
the field theory result of~\cite{Bern:1997nh} (this means that we just
have double-box and non-planar double-box graphs).

In terms of the string loop-integrand
$I^{(2)}\simeq I^{(2)}(z_i,\ell_i)$ of $A^{(2)}_\mathrm{orient}$, the
monodromy relations of \cite{Tourkine:2016bak} at two loops read:
\begin{equation}
k_1\cdot k_2\,I^{(2)}(2134)
+k_1\cdot (k_2+k_3)I^{(2)}(2314)
-\ell_1\cdot k_1 I^{(2)}(234|1|.)-\ell_2\cdot k_1 I^{(2)}(234|.|1)
\simeq0\,.
  \label{eq:2loop-monodromy}
\end{equation}
The two terms on the rightmost part correspond to non-planar
amplitudes where the particle $1$ is integrated along the first and
second inner disks of the two-loop open string graph, respectively
(from left to right in fig.~\ref{fig:two-loop-ex}). The $\simeq$
symbol means ``up to terms that vanish after momentum
integration''. These are generated by integrals along the boundary of
the cut surface and correspond to loop momentum shifts.

\begin{equation}
  \label{eq:test}
  \includegraphics[]{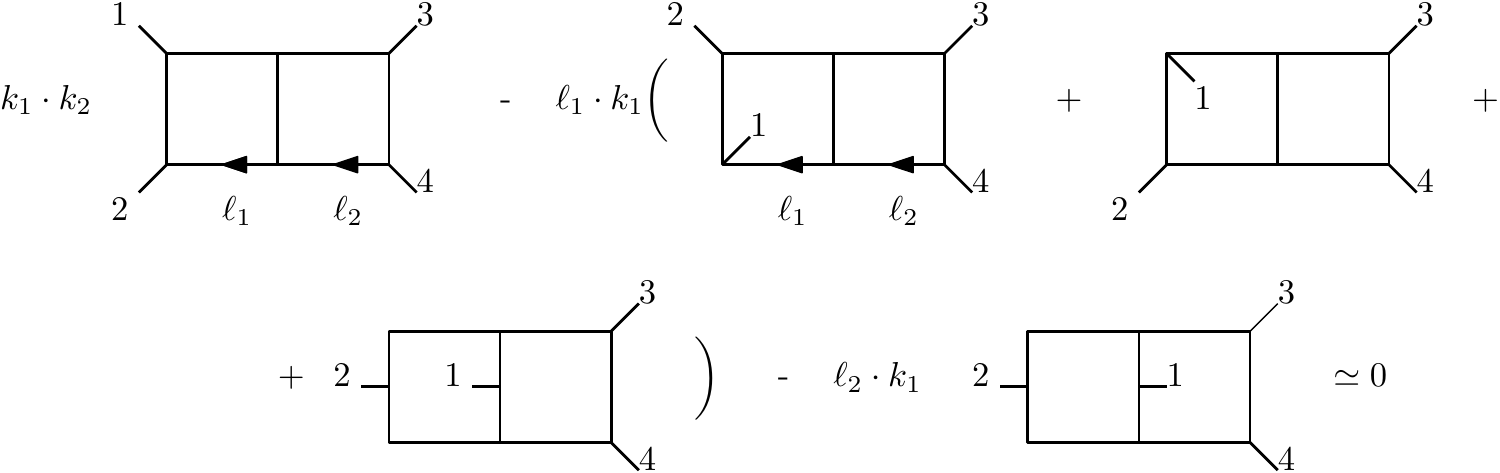}
\end{equation}
All other diagrams are suppressed by supersymmetry in the field theory
limit, because of the properties of the tropical limit of the
integrand given by $\mathcal{Y}_S$ that we just described. These
graphs are scalar graphs, with their denominator, and with the same
numerator $s_{12}=-(k_1+k_2)^2$ because $\mathcal{Y}_s=s_{34}=s_{12}$
(again see \cite[Tab.~1, p.~41]{Tourkine:2013rda}).
Therefore, these graphs are just scalar graphs with constant
numerator.

Using the antisymmetry of the three-point vertex
\cite{Tourkine:2016bak,Ochirov:2017jby}, we can equate the two graphs
on the second line up to a sign and reduce the factors in front of the
graphs to differences of propagators,
\begin{align}
  & \ell_1\cdot k_1 = (\ell_1+k_1)^2-\ell_1^2,\\
  & (\ell_1+k_2)\cdot  k_1 =
    (\ell_1+k_ 1+k_2)^2-(\ell_1+k_2)^2,\\
  &(-\ell_1+\ell_2)\cdot k_1 = (\ell_1-\ell _2)^2-(\ell _1-\ell _2-k_1)^2\,.
\end{align}
In this way, six terms are
produced which almost cancel pairwise:
\begin{equation}
  \label{eq:contracted}
    \includegraphics[]{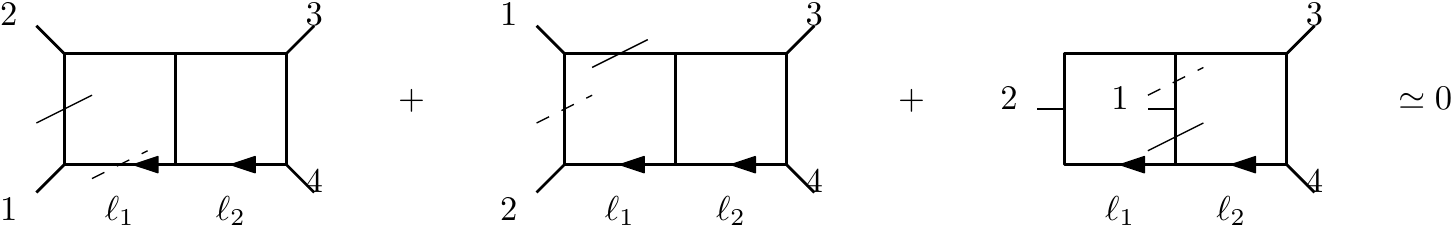}
\end{equation}
In this equation, the plain (resp. dashed) lines correspond to a
positive (resp. negative sign). Four terms cancel pairwise, while two,
the negative contribution of the first graph and the positive one of
the last graph, differ by a shift in the loop momentum as:
\begin{equation}
  \label{eq:almost-cancel}
  \includegraphics[]{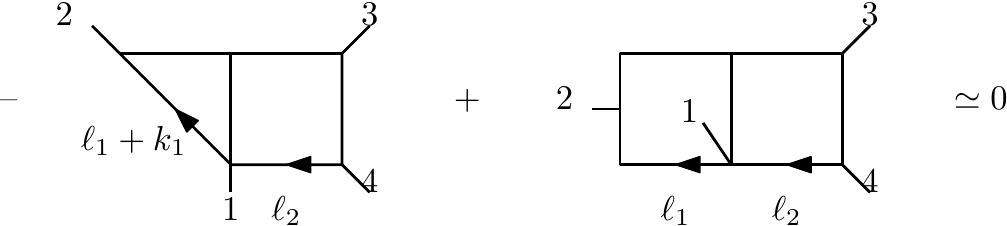}
\end{equation}
Because the relation is exact at fixed loop momentum, this gives a
precise definition the terms in the right-hand side. A more graphical
explanation of this phenomenon can be found in
\cite{Ochirov:2017jby}.
At any rate, after loop momentum integration, these terms cancel, as
they should. Note that for more generic amplitudes, the numerators are
not simply constants anymore and the field theory limit of the terms
on the right-hand side could provide interesting physical
quantities. These will be studied elsewhere.

This derivation provides a stronger check than the unitarity cut check
that was originally performed in \cite{Tourkine:2016bak}.

\subsection{A relation at three loops}
\label{sec:relation-at-three}

What we have seen so far is that the string representation in terms of
loop momentum induces a global definition of the loop momentum. Now we
will investigate a new phenomenon related to this that arises at three
loops: there are two different vacuum topologies of
1-particle-irreducible graphs, mercedes and ladder, \textit{which
  share the same loop momentum}.

After characterizing this effect, we will work out an example of
application of the monodromy relations to support further that their
connection to the BCJ color-kinematics duality extends to all loops.

\begin{figure}[t]
  \centering
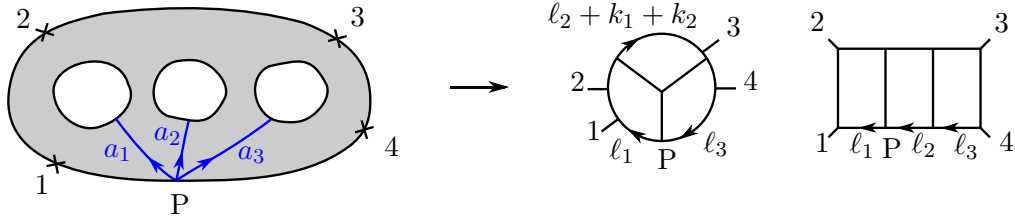
  \caption{The two graphs on the right-hand side are generated by the
    field theory limit of the open string graph on the left-hand
    side. It is an interesting exercise for the reader to work out
    this example explicitly.}
  \label{fig:3-loop}
\end{figure}

Figure \ref{fig:3-loop} displays two representative graphs that follow
from the field theory limit of the open string graph on the left-hand
side. Both these graphs appear under the same loop-momentum integral
in the field theory limit, and provide a natural correspondence
between the same loop momentum but in different graphs.

The monodromy relations in string theory at three-loops are obtained
by followed the method exposed in \cite{Tourkine:2016bak}. Circulating
the leg $1$ inside a previously planar graph with ordering $1<2<3<4$
as in fig.~\ref{fig:3-loop} yields a relation, whose field theory
limit is given by
\begin{multline}
k_1\cdot k_2\,I^{(3)}(2134)
+k_1\cdot (k_2+k_3)I^{(3)}(2314)\cr
- \ell_1\cdot k_1 I^{(3)}(234|1|.|.)
-\ell_2\cdot k_1 I^{(3)}(234|.|1|.)
-\ell_3\cdot k_1  I^{(3)}(234|.|.|1)\simeq 0\,.
  \label{eq:3loop-monodromy}
\end{multline}
The notations are similar to those of eq.~(\ref{eq:2loop-monodromy}).
The terms on the second line correspond to non-planar amplitudes where
the particle $1$ is integrated along the first, second and third inner
disks, respectively, according to the numbering of the $a$ cycles in
fig.~\ref{fig:3-loop}.

Many graphs arise in this integrand relation.\footnote{Counting by
  hand graphs with no-triangles (having in mind N=4 super-Yang-Mills)
  give an $\leq O(150)$ graphs.}  They mix different topologies and
orderings. The systematics of the propagator cancelation is similar to
what happens at two loops. We illustrate this below for a particular
subset of these graphs, which will give stronger evidence that a BCJ
representation always satisfy the monodromy relation, up to the
loop-momentum shifting terms.

The main point is that this sum of graphs can be re-organized into BCJ
triplets. For instance, the following four terms appear in the sum in
the left-hand side of (\ref{eq:3loop-monodromy}):
\begin{multline}
\includegraphics{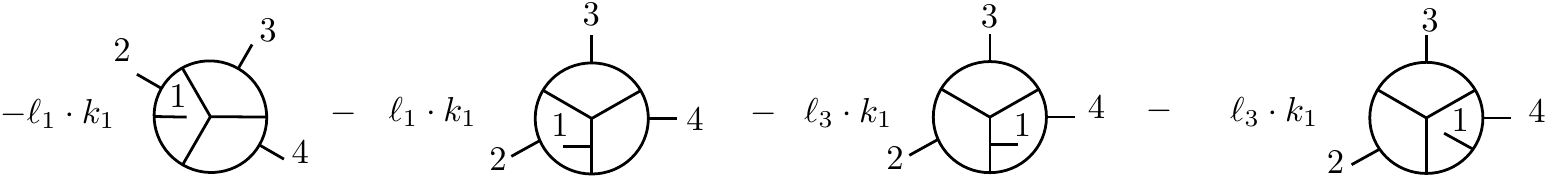}
  \\+\ldots\simeq0
%
  \label{fig:BCJ-3loop}
\end{multline}
where $+...$ indicate the rest of the terms of the sum. These graphs
represent full integrands: numerators over denominator. They are those
of \textit{any} gauge theory we started with in the open
string.\footnote{One strength of the monodromy relations is that they
  are universal.}

With the momenta are distributed as in fig~\ref{fig:3-loop}, it is
easy to see that the factors in front of the graphs indeed recombine
into differences of irreducible propagators which organise themselves
as a BCJ identity with shifted momenta of the form:
\begin{multline}
  \label{eq:bcj-3loo-shifted}
-D\bigg(\vcenter{\hbox{\includegraphics{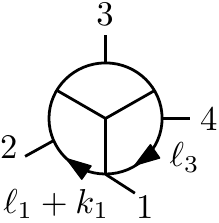}}}\bigg)\times
N\bigg(\vcenter{\hbox{\includegraphics{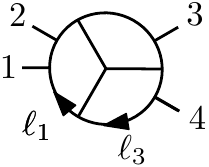}}}\bigg)+\\
D\bigg(\vcenter{\hbox{\includegraphics{./graphs/den-shifted.pdf}}}\bigg)\times
\bigg\{N\bigg(\vcenter{\hbox{\includegraphics{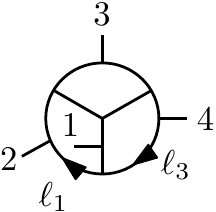}}}\bigg)+
N\bigg(\vcenter{\hbox{\includegraphics{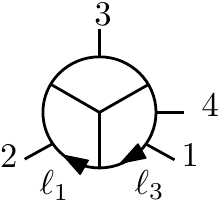}}}\bigg)\bigg\}\,.
\end{multline}
Here, $D(\cdot)$ is the scalar denominator corresponding to the graph
(non-obvious loop-momentum locations are depicted, $\ell_2$ is not
affected) and $N(\cdot)$ is the numerator of the corresponding
graph. I have used again the antisymmetry of the three-point vertex
\begin{equation}
  \label{eq:antisym}
  N\bigg(\vcenter{\hbox{\includegraphics{./graphs/gr2.pdf}}}\bigg) = -
  N\bigg(\vcenter{\hbox{\includegraphics{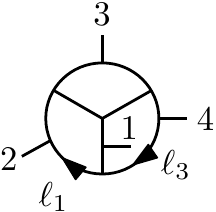}}}\bigg)\,.
\end{equation}
The three terms above therefore combine into a BCJ triplet involving
some loop momentum shifts on top of denominators with one propagator
canceled.

We have worked out the specific case that is the most delicate,
i.e. the one that involves loop momentum shifts. The other triplet
identities are simpler and therefore it is very reasonable to guess
that the property persists for all types of tri-valent graphs also
including those with internal triangles, bubbles or not -- see
\cite{Ochirov:2017jby} for examples at one loop.

Note that an identity that would mix up mercedes and ladder topologies
requires to apply the monodromy relations twice or to start from a
non-planar amplitude.

To sum up the relations that stem from the monodromy relations in the
field theory limit, we schematically denote by the letter $J_{G,e}$
the sum of BCJ triplets for the graph $G$ with one inner edge $e$
contracted. We obtain:
\begin{equation}
  \sum_{G,e} \frac{J_{G,e}}{D_{G,e}}\simeq 0
\end{equation}
In the generalised double copy construction~\cite{Bern:2017ucb}, these
$J$-functions generate higher point vertices that need to be canceled
by introducing contact terms. The structure of these objects is still
quite poorly understood, in particular how to simplify them as much
as possible, and it would be interesting to see if the
monodromy relations can provide some formal constraints on these
objects, maybe in relation to the loop shifting terms of the
right-hand side.

\section{Discussion}
\label{sec:discussion}

\paragraph{An integrand in field theory.}

In this paper we analyzed some aspects related to the definition of
the loop momentum in string and field theory. This formalism was
mostly developed to be applied to the monodromy relations, but it
would be very interesting to see if the global integrand defined in
this way has any nice physical properties.

Furthermore, in standard perturbative field theory there is no
particular notion of field theory integrand except in the case of
planar amplitudes: this has lead to remarkable constructions such as
the all-loop integrand for planar $\mathcal{N}=4$ super-Yang Mills
\cite{ArkaniHamed:2010kv} and the amplituhedron
\cite{Arkani-Hamed:2013jha}. This program was then extended to
gravitational theories in \cite{Herrmann:2016qea} and it would be
interesting to see if all these constructions are connected to the
general considerations presented in this paper.

The ambitwistor string~\cite{Mason:2013sva}, based on the scattering
equation formalism \cite{Cachazo:2013iea} also provide loop integrands
\cite{Adamo:2013tsa,Ohmori:2015sha,Casali:2014hfa,Geyer:2015bja,Geyer:2015jch,Cachazo:2015aol,He:2015yua,He:2015wgf,Geyer:2016wjx,Geyer:2018xwu}. The
bottleneck in pushing this formalism to all loops has so far been the
understanding of the geometry of the moduli space and the connection
to the zero-modes (loop-momentum) in the path integral. There is no
doubts that a better understanding of the loop momentum in string
theory should help to fix these issues.

\paragraph{Kawai-Lewellen-Tye}

Since they realize splitting of the holomorphic and anti-holomorphic
degrees of freedom in string integrands, loop momentum representations
should also be linked to the extension of the tree-level
Kawai-Lewellen-Tye \cite{Kawai:1985xq,BjerrumBohr:2010hn} formulae to
loop-level. Recently, Mizera has reformulated in the language of
twisted cycles this program~\cite{Mizera:2017cqs,Mizera:2017rqa} and a
deeper understanding of the loop momentum will be necessary to
understand these constructions at loop-level where global monodromies
arise.
Relatedly, it would be very interesting to see if these relations can
be extended to amplitudes relations. The theory developped in
\cite{Mastrolia:2018uzb} for field theory integrands, inspired from
\cite{Mizera:2017cqs}, would seem like a natural starting point to
study these questions.
Relatedly, ``generalized elliptic functions'' have been introduced in
\cite{Mafra:2017ioj,Mafra:2018nla,Mafra:2018qqe,Mafra:2018pll} and it
would be interesting to see if a proper treatment of the loop momentum
can help in characterizing the nature of these objects.

\paragraph{Twisted strings and modular invariance}

Twisted strings\footnote{Also called ``left-handed''
  \cite{Siegel:2015axg} or ``chiral-strings''
  \cite{Huang:2016bdd,Leite:2016fno}. The ``twisted string''
  terminology was used in \cite{Casali:2017mss} because toroidal
  compactifications allows what are chiral strings in flat space to
  acquire non-trivial excitations, hence they are not really chiral.}
are the tensionful versions of ambitwistor strings. To my knowledge,
the first time such a construction was mentioned is in the paper
\cite{Hwang:1998gs}, and in spirit they were present in
\cite{Gamboa:1989zc,Gamboa:1989px} already. Classically, they are just
identical to traditional string theory; but their quantization is
modified (different operator ordering) which results in a truncated
spectrum. The cleanest way to understrand their scattering amplitudes
is at tree-level so far, via the twisted period relations of
\cite{Mizera:2017rqa}.

It is conjectured \cite{Siegel:2015axg} that the loop level version
should also involve only a change in oscillator modes of the string,
therefore all we said here about the loop-momentum zero-modes should
apply to the twisted string too.
However, loop amplitudes have proven difficult to write so far, and a
very good hope to guess them would be to generalize the twisted period
relations to loop-level. This ties in with the previous paragraph on
KLT.

One could even think of using these twisted string loop amplitudes to
then take the ambitwistor string limit (tensionless limit of the
twisted string, see \cite{Siegel:2015axg,Leite:2016fno,Huang:2016bdd}
and \cite{Casali:2017zkz,Casali:2016atr}).
But one may doubt that this could produce a sensible answer, mostly
because the loop momentum breaks modular invariance
\cite{DHoker:1988pdl} and the saddle point equations of the
tensionless limit~\cite{Gross:1987ar} seem to induce a maximum value
for the loop momentum, while all values should be allowed and
integrated on to give back the original integral without loop
momentum. This problem will be studied elsewhere.

\paragraph{Acknowledgements}
I would like to thank David Andriot, Enrico Hermann, Alexander
Ochirov, Julio Parra-Martinez, Boris
Pioline, Amit Sever and Sasha Zhiboedov for
useful discussions and comments. This research project has been
supported by a Marie Skłodowska-Curie Individual Fellowship of the
European Commission’s Horizon 2020 Programme under contract number
749793 NewLoops.

\appendix

\section{Definitions}
\label{sec:definitions}


Here are the conventions that are used in this paper (we follow mostly
Kiritsis \cite{Kiritsis:2007zza})
\begin{align}
z=\sigma^1+i \sigma^2,\quad \bar
  z= \sigma^1-i\sigma^2,\quad
  \frac{\partial}{\partial z} = \frac 12(\partial_1-i
\partial_2),\quad   \frac{\partial}{\partial \bar z} =\frac12 (\partial_1 + i \partial_2)
\end{align}
The metric reads $d^2 s = 2g_{z \bar z}d z d\bar z$, therefore
$\sqrt{g}=g_{z \bar z}$ such that the
volume measure is given by $\int \sqrt{g_{ab}} d^2 \sigma = \int
\sqrt{g} d^2 z$ which yields
\begin{equation}
  2d\sigma_1 d \sigma_2= d^2 z = i dz \wedge d\bar z
\end{equation}
For the diagonal metric $g_{ab}=\mathrm{diag}(1,1)$ we have $g_{z \bar z}=1/2$
which implies that $\delta^2(z,\bar
z)=\frac{1}{2\sqrt{g}}\delta(x)\delta(y)=\delta(x)\delta(y)$, as this yields
\begin{equation}
  \int \sqrt{g} d^2 z \delta^2(z,\bar z)=1
\end{equation}
We also
have
\begin{equation}
  \label{eq:delta-def-appendix}
  \partial_{\bar z} \frac{1}{z} = \partial_z \frac{1}{\bar z}=2\pi \delta^2(z,\bar z) 
\end{equation}


We will use the language of differential forms (all conventions
are spelled out in appendix~\ref{sec:definitions}),
upon which,
essentially,
c\begin{equation}
  d^2z = i{dz\wedge d\bar z},\quad \partial = \d dz,\quad \bar
  \partial = \db d\bar z\,\quad d =\partial+\bar \partial
  \label{eq:norm}
\end{equation}
where $d$ is the standard differential operator, $d^2=0$. The $i$
normalisation factor will be important soon. Stokes theorem states
that, for $\omega$ a $k$-form and $D$ a $(k+1)$-chain, we have
\begin{equation}
  \label{eq:stokes}
  \int_{\partial D} \omega = \int_D d\omega
\end{equation}
where $\partial D$ is the boundary of $D$.



The prime form is a $(-1/2,0)\otimes(-1/2,0)$ bi-holomorphic form
defined on the universal covering of the surface by
\begin{equation}
  E(x,y)=
\frac{\theta[ \nu](\int_x^y
(\omega_1,...,\omega_g)|{\Omega})}{h_{ \nu}(x) h_{ \nu}(y)}\,\in 
\mathbb{C}\,,
\label{e:primeform}
\end{equation}
where $h_{ \nu}(x)^2 = \sum_I \omega_I \partial_I \theta[ \nu]
(0|\Omega)$ are half-differentials (section of the square-root of the
canonical bundle). It is independent of the spin structure chosen
to define it.

The Riemann theta functions are defined by
\begin{equation}
  \theta[ \nu] ( \zeta| \Omega) =
   \sum_{ n\in\mathbb{Z}^g} e^{i\pi ({n}+{\beta})\cdot  \Omega ({n}+{\beta})} e^{2 i
      \pi ({n}+{\beta})\cdot ({\zeta}+{\alpha})}
    \label{e:thetachar}
\end{equation}
where $\ab= \nu \in (\mathbb{Z}/\mathbb{Z}_2)^{2g}$ is a theta
characteristic. They have monodromies that can be found in standard
textbooks, which lead to the following monodromies for the prime form \cite{DHoker:1988pdl};
\begin{equation}
E(x,y)\to\exp(- \Omega_{JJ}/2 - \int_x^y \omega_J) E(x,y)
\label{e:multE}\,.
\end{equation}
and trivial signs along $a$ cycles.



\bibliography{biblio.bib}
\bibliographystyle{JHEP}

\end{document}

%% file: decomposition.pdf_tex
\begingroup%
  \makeatletter%
  \providecommand\color[2][]{%
    \errmessage{(Inkscape) Color is used for the text in Inkscape, but the package 'color.sty' is not loaded}%
    \renewcommand\color[2][]{}%
  }%
  \providecommand\transparent[1]{%
    \errmessage{(Inkscape) Transparency is used (non-zero) for the text in Inkscape, but the package 'transparent.sty' is not loaded}%
    \renewcommand\transparent[1]{}%
  }%
  \providecommand\rotatebox[2]{#2}%
  \ifx\svgwidth\undefined%
    \setlength{\unitlength}{372.83367153bp}%
    \ifx\svgscale\undefined%
      \relax%
    \else%
      \setlength{\unitlength}{\unitlength * \real{\svgscale}}%
    \fi%
  \else%
    \setlength{\unitlength}{\svgwidth}%
  \fi%
  \global\let\svgwidth\undefined%
  \global\let\svgscale\undefined%
  \makeatother%
  \begin{picture}(1,0.26146949)%
    \put(0,0){\includegraphics[width=\unitlength,page=1]{decomposition.pdf}}%
    \put(0.31373211,0.07334142){\color[rgb]{0,0,0}\makebox(0,0)[lb]{\smash{$\color{red}a_2$}}}%
    \put(0.17202206,0.06713653){\color[rgb]{0,0,0}\makebox(0,0)[lb]{\smash{$\color{red}a_2$}}}%
    \put(0,0){\includegraphics[width=\unitlength,page=2]{decomposition.pdf}}%
    \put(0.09720005,0.13806517){\color[rgb]{0,0,0}\makebox(0,0)[lb]{\smash{$\color{blue}b_1$}}}%
    \put(0.34193422,0.20976421){\color[rgb]{0,0,0}\makebox(0,0)[lb]{\smash{$\color{blue}b_2$}}}%
    \put(0,0){\includegraphics[width=\unitlength,page=3]{decomposition.pdf}}%
    \put(0.84151354,0.0049292){\color[rgb]{0,0,0}\makebox(0,0)[lb]{\smash{$\color{red}a_2$}}}%
    \put(0.93081437,0.17882101){\color[rgb]{0,0,0}\makebox(0,0)[lb]{\smash{$\color{red}a_2^{-1}$}}}%
    \put(0.92966906,0.07994626){\color[rgb]{0,0,0}\makebox(0,0)[lb]{\smash{$\color{blue}b_2$}}}%
    \put(0.88139111,0.23846515){\color[rgb]{0,0,0}\makebox(0,0)[lb]{\smash{$\color{blue}b_2^{-1}$}}}%
    \put(0.77573962,0.24407871){\color[rgb]{0,0,0}\makebox(0,0)[lb]{\smash{$\color{red}a_1$}}}%
    \put(0.68235969,0.09110906){\color[rgb]{0,0,0}\makebox(0,0)[lb]{\smash{$\color{red}a_1^{-1}$}}}%
    \put(0.69914183,0.18277285){\color[rgb]{0,0,0}\makebox(0,0)[lb]{\smash{$\color{blue}b_1$}}}%
    \put(0.73415709,0.02381853){\color[rgb]{0,0,0}\makebox(0,0)[lb]{\smash{$\color{blue}b_1^{-1}$}}}%
  \end{picture}%
\endgroup%

%% file: homology_example_workedout.pdf_tex
\begingroup%
  \makeatletter%
  \providecommand\color[2][]{%
    \errmessage{(Inkscape) Color is used for the text in Inkscape, but the package 'color.sty' is not loaded}%
    \renewcommand\color[2][]{}%
  }%
  \providecommand\transparent[1]{%
    \errmessage{(Inkscape) Transparency is used (non-zero) for the text in Inkscape, but the package 'transparent.sty' is not loaded}%
    \renewcommand\transparent[1]{}%
  }%
  \providecommand\rotatebox[2]{#2}%
  \ifx\svgwidth\undefined%
    \setlength{\unitlength}{358.95428522bp}%
    \ifx\svgscale\undefined%
      \relax%
    \else%
      \setlength{\unitlength}{\unitlength * \real{\svgscale}}%
    \fi%
  \else%
    \setlength{\unitlength}{\svgwidth}%
  \fi%
  \global\let\svgwidth\undefined%
  \global\let\svgscale\undefined%
  \makeatother%
  \begin{picture}(1,0.6246083)%
    \put(0,0){\includegraphics[width=\unitlength,page=1]{homology_example_workedout.pdf}}%
    \put(0.14774502,0.13590053){\color[rgb]{0,0,0}\makebox(0,0)[lb]{\smash{$z_1$}}}%
    \put(0.22662101,0.44954589){\color[rgb]{0,0,0}\makebox(0,0)[lb]{\smash{$z_2$}}}%
    \put(0.6317977,0.56201262){\color[rgb]{0,0,0}\makebox(0,0)[lb]{\smash{$z_3$}}}%
    \put(0.78881547,0.47780979){\color[rgb]{0,0,0}\makebox(0,0)[lb]{\smash{$z_4$}}}%
    \put(0.89506819,0.26347533){\color[rgb]{0,0,0}\makebox(0,0)[lb]{\smash{$z_5$}}}%
    \put(0,0){\includegraphics[width=\unitlength,page=2]{homology_example_workedout.pdf}}%
    \put(0.5452283,0.38096976){\color[rgb]{0,0,0}\makebox(0,0)[lb]{\smash{$\color{red}a_2$}}}%
    \put(0.76561534,0.15030278){\color[rgb]{0,0,0}\makebox(0,0)[lb]{\smash{$\color{red}a_3$}}}%
    \put(0.48428791,0.22524873){\color[rgb]{0,0,0}\makebox(0,0)[lb]{\smash{$\color{red}a_1$}}}%
    \put(0.45715956,0.57979534){\color[rgb]{0,0,0}\makebox(0,0)[lb]{\smash{$C$}}}%
    \put(0,0){\includegraphics[width=\unitlength,page=3]{homology_example_workedout.pdf}}%
    \put(0.23333745,0.12377058){\color[rgb]{0,0,0}\makebox(0,0)[lb]{\smash{${b_1}^{-1}$}}}%
    \put(0.14892562,0.33963599){\color[rgb]{0,0,0}\makebox(0,0)[lb]{\smash{$b_1$}}}%
    \put(0,0){\includegraphics[width=\unitlength,page=4]{homology_example_workedout.pdf}}%
    \put(0.14108628,0.42165902){\color[rgb]{0,0,0}\makebox(0,0)[lb]{\smash{$c_2$}}}%
    \put(0.09760875,0.11800997){\color[rgb]{0,0,0}\makebox(0,0)[lb]{\smash{$c_1$}}}%
  \end{picture}%
\endgroup%

%% file: plumbing.pdf_tex
\begingroup%
  \makeatletter%
  \providecommand\color[2][]{%
    \errmessage{(Inkscape) Color is used for the text in Inkscape, but the package 'color.sty' is not loaded}%
    \renewcommand\color[2][]{}%
  }%
  \providecommand\transparent[1]{%
    \errmessage{(Inkscape) Transparency is used (non-zero) for the text in Inkscape, but the package 'transparent.sty' is not loaded}%
    \renewcommand\transparent[1]{}%
  }%
  \providecommand\rotatebox[2]{#2}%
  \ifx\svgwidth\undefined%
    \setlength{\unitlength}{512.77872479bp}%
    \ifx\svgscale\undefined%
      \relax%
    \else%
      \setlength{\unitlength}{\unitlength * \real{\svgscale}}%
    \fi%
  \else%
    \setlength{\unitlength}{\svgwidth}%
  \fi%
  \global\let\svgwidth\undefined%
  \global\let\svgscale\undefined%
  \makeatother%
  \begin{picture}(1,0.29978812)%
    \put(0,0){\includegraphics[width=\unitlength,page=1]{plumbing.pdf}}%
    \put(0.76856223,0.12616543){\color[rgb]{0,0,0}\makebox(0,0)[lb]{\smash{...}}}%
    \put(0,0){\includegraphics[width=\unitlength,page=2]{plumbing.pdf}}%
    \put(0.16200499,0.15078437){\color[rgb]{0,0,0}\makebox(0,0)[lb]{\smash{$p_a$}}}%
    \put(0.46098745,0.17150087){\color[rgb]{0,0,0}\makebox(0,0)[lb]{\smash{$p_b$}}}%
    \put(0.06295213,0.21074256){\color[rgb]{0,0,0}\makebox(0,0)[lb]{\smash{$C_a''$}}}%
    \put(0.23806272,0.13102888){\color[rgb]{0,0,0}\makebox(0,0)[lb]{\smash{$C_a'$}}}%
    \put(0.36431873,0.23163614){\color[rgb]{0,0,0}\makebox(0,0)[lb]{\smash{$C_b'$}}}%
    \put(0.54463013,0.2074974){\color[rgb]{0,0,0}\makebox(0,0)[lb]{\smash{$C_b''$}}}%
    \put(0,0){\includegraphics[width=\unitlength,page=3]{plumbing.pdf}}%
    \put(0.67854147,0.10497572){\color[rgb]{0,0,0}\makebox(0,0)[lb]{\smash{$\color{red}a_1$}}}%
    \put(0.78644541,0.06014486){\color[rgb]{0,0,0}\makebox(0,0)[lb]{\smash{$\color{red}a_{g-1}$}}}%
    \put(0,0){\includegraphics[width=\unitlength,page=4]{plumbing.pdf}}%
    \put(0.36556427,0.07674264){\color[rgb]{0,0,0}\makebox(0,0)[lb]{\smash{$\color{red}a_{g}$}}}%
    \put(0,0){\includegraphics[width=\unitlength,page=5]{plumbing.pdf}}%
    \put(0.25866945,0.21474116){\color[rgb]{0,0,0}\makebox(0,0)[lb]{\smash{$\color{blue}b_{g}$}}}%
    \put(0,0){\includegraphics[width=\unitlength,page=6]{plumbing.pdf}}%
    \put(0.19254173,0.21064364){\color[rgb]{0,0,0}\makebox(0,0)[lb]{\smash{$z_a$}}}%
    \put(0.41421328,0.15426942){\color[rgb]{0,0,0}\makebox(0,0)[lb]{\smash{$z_b$}}}%
  \end{picture}%
\endgroup%

%% file: plumbing-pinch.pdf_tex
\begingroup%
  \makeatletter%
  \providecommand\color[2][]{%
    \errmessage{(Inkscape) Color is used for the text in Inkscape, but the package 'color.sty' is not loaded}%
    \renewcommand\color[2][]{}%
  }%
  \providecommand\transparent[1]{%
    \errmessage{(Inkscape) Transparency is used (non-zero) for the text in Inkscape, but the package 'transparent.sty' is not loaded}%
    \renewcommand\transparent[1]{}%
  }%
  \providecommand\rotatebox[2]{#2}%
  \ifx\svgwidth\undefined%
    \setlength{\unitlength}{512.77872479bp}%
    \ifx\svgscale\undefined%
      \relax%
    \else%
      \setlength{\unitlength}{\unitlength * \real{\svgscale}}%
    \fi%
  \else%
    \setlength{\unitlength}{\svgwidth}%
  \fi%
  \global\let\svgwidth\undefined%
  \global\let\svgscale\undefined%
  \makeatother%
  \begin{picture}(1,0.29978812)%
    \put(0,0){\includegraphics[width=\unitlength,page=1]{plumbing-pinch.pdf}}%
    \put(0.76856223,0.12773253){\color[rgb]{0,0,0}\makebox(0,0)[lb]{\smash{...}}}%
    \put(0,0){\includegraphics[width=\unitlength,page=2]{plumbing-pinch.pdf}}%
    \put(0.19254173,0.21221073){\color[rgb]{0,0,0}\makebox(0,0)[lb]{\smash{$z_a$}}}%
    \put(0.40637782,0.19919273){\color[rgb]{0,0,0}\makebox(0,0)[lb]{\smash{$z_b$}}}%
    \put(0,0){\includegraphics[width=\unitlength,page=3]{plumbing-pinch.pdf}}%
    \put(0.56188982,0.24829125){\color[rgb]{0,0,0}\makebox(0,0)[lb]{\smash{$C$}}}%
    \put(0,0){\includegraphics[width=\unitlength,page=4]{plumbing-pinch.pdf}}%
    \put(0.58816549,0.15746329){\color[rgb]{0,0,0}\makebox(0,0)[lb]{\smash{$P$}}}%
    \put(0.30257887,0.06687066){\color[rgb]{0,0,0}\makebox(0,0)[lb]{\smash{$z_i$}}}%
    \put(0,0){\includegraphics[width=\unitlength,page=5]{plumbing-pinch.pdf}}%
    \put(0.22292227,0.13014031){\color[rgb]{0,0,0}\makebox(0,0)[lb]{\smash{$y_a$}}}%
    \put(0,0){\includegraphics[width=\unitlength,page=6]{plumbing-pinch.pdf}}%
  \end{picture}%
\endgroup%

%% file: loops.pdf_tex
\begingroup%
  \makeatletter%
  \providecommand\color[2][]{%
    \errmessage{(Inkscape) Color is used for the text in Inkscape, but the package 'color.sty' is not loaded}%
    \renewcommand\color[2][]{}%
  }%
  \providecommand\transparent[1]{%
    \errmessage{(Inkscape) Transparency is used (non-zero) for the text in Inkscape, but the package 'transparent.sty' is not loaded}%
    \renewcommand\transparent[1]{}%
  }%
  \providecommand\rotatebox[2]{#2}%
  \ifx\svgwidth\undefined%
    \setlength{\unitlength}{482.26918495bp}%
    \ifx\svgscale\undefined%
      \relax%
    \else%
      \setlength{\unitlength}{\unitlength * \real{\svgscale}}%
    \fi%
  \else%
    \setlength{\unitlength}{\svgwidth}%
  \fi%
  \global\let\svgwidth\undefined%
  \global\let\svgscale\undefined%
  \makeatother%
  \begin{picture}(1,0.15180366)%
    \put(0,0){\includegraphics[width=\unitlength,page=1]{loops.pdf}}%
    \put(0.14218606,0.05845755){\color[rgb]{0,0,0}\makebox(0,0)[lb]{\smash{$\color{red} a_g$}}}%
    \put(0.28267556,-0.05825197){\color[rgb]{0,0,0}\makebox(0,0)[lb]{\smash{}}}%
    \put(0,0){\includegraphics[width=\unitlength,page=2]{loops.pdf}}%
    \put(0.23997107,0.06250689){\color[rgb]{0,0,0}\makebox(0,0)[lb]{\smash{...}}}%
    \put(0,0){\includegraphics[width=\unitlength,page=3]{loops.pdf}}%
    \put(0.52057625,0.06951149){\color[rgb]{0,0,0}\makebox(0,0)[lb]{\smash{$\color{red} a_g$}}}%
    \put(0,0){\includegraphics[width=\unitlength,page=4]{loops.pdf}}%
    \put(0.61836119,0.07356083){\color[rgb]{0,0,0}\makebox(0,0)[lb]{\smash{...}}}%
    \put(0,0){\includegraphics[width=\unitlength,page=5]{loops.pdf}}%
    \put(0.94035453,0.06588872){\color[rgb]{0,0,0}\makebox(0,0)[lb]{\smash{...}}}%
    \put(0.91156133,0.03875096){\color[rgb]{0,0,0}\makebox(0,0)[lb]{\smash{$\color{red} a_g$}}}%
    \put(0,0){\includegraphics[width=\unitlength,page=6]{loops.pdf}}%
    \put(0.06854264,0.06780882){\color[rgb]{0,0,0}\makebox(0,0)[lb]{\smash{$C$}}}%
  \end{picture}%
\endgroup%

%% file: 2loop-example.pdf_tex
\begingroup%
  \makeatletter%
  \providecommand\color[2][]{%
    \errmessage{(Inkscape) Color is used for the text in Inkscape, but the package 'color.sty' is not loaded}%
    \renewcommand\color[2][]{}%
  }%
  \providecommand\transparent[1]{%
    \errmessage{(Inkscape) Transparency is used (non-zero) for the text in Inkscape, but the package 'transparent.sty' is not loaded}%
    \renewcommand\transparent[1]{}%
  }%
  \providecommand\rotatebox[2]{#2}%
  \ifx\svgwidth\undefined%
    \setlength{\unitlength}{434.01810309bp}%
    \ifx\svgscale\undefined%
      \relax%
    \else%
      \setlength{\unitlength}{\unitlength * \real{\svgscale}}%
    \fi%
  \else%
    \setlength{\unitlength}{\svgwidth}%
  \fi%
  \global\let\svgwidth\undefined%
  \global\let\svgscale\undefined%
  \makeatother%
  \begin{picture}(1,0.15738051)%
    \put(0,0){\includegraphics[width=\unitlength,page=1]{2loop-example.pdf}}%
    \put(0.84240251,0.00262396){\color[rgb]{0,0,0}\makebox(0,0)[lb]{\smash{$\ell_1$}}}%
    \put(0.72933829,0.08088671){\color[rgb]{0,0,0}\makebox(0,0)[lb]{\smash{$\ell_1+k_1$}}}%
    \put(0,0){\includegraphics[width=\unitlength,page=2]{2loop-example.pdf}}%
    \put(0.91035173,0.00262396){\color[rgb]{0,0,0}\makebox(0,0)[lb]{\smash{$\ell_2$}}}%
    \put(0,0){\includegraphics[width=\unitlength,page=3]{2loop-example.pdf}}%
    \put(0.8876907,0.05616865){\color[rgb]{0,0,0}\makebox(0,0)[lb]{\smash{$\ell_1+\ell_2$}}}%
    \put(0,0){\includegraphics[width=\unitlength,page=4]{2loop-example.pdf}}%
    \put(0.02984499,0.00488728){\color[rgb]{0,0,0}\makebox(0,0)[lb]{\smash{1}}}%
    \put(0.02490249,0.13901612){\color[rgb]{0,0,0}\makebox(0,0)[lb]{\smash{2}}}%
    \put(0,0){\includegraphics[width=\unitlength,page=5]{2loop-example.pdf}}%
    \put(0.089527,0.07021808){\color[rgb]{0,0,0}\makebox(0,0)[lb]{\smash{3}}}%
    \put(0.25945422,0.07554359){\color[rgb]{0,0,0}\makebox(0,0)[lb]{\smash{4}}}%
    \put(0,0){\includegraphics[width=\unitlength,page=6]{2loop-example.pdf}}%
    \put(0.39158009,0.00797684){\color[rgb]{0,0,0}\makebox(0,0)[lb]{\smash{1}}}%
    \put(0.39404019,0.14157695){\color[rgb]{0,0,0}\makebox(0,0)[lb]{\smash{2}}}%
    \put(0.49592303,0.08469511){\color[rgb]{0,0,0}\makebox(0,0)[lb]{\smash{3}}}%
    \put(0.656979,0.07881461){\color[rgb]{0,0,0}\makebox(0,0)[lb]{\smash{4}}}%
    \put(0,0){\includegraphics[width=\unitlength,page=7]{2loop-example.pdf}}%
    \put(0.78719772,0.01320228){\color[rgb]{0,0,0}\makebox(0,0)[lb]{\smash{1}}}%
    \put(0.7891195,0.14680239){\color[rgb]{0,0,0}\makebox(0,0)[lb]{\smash{2}}}%
    \put(0.84797303,0.08992053){\color[rgb]{0,0,0}\makebox(0,0)[lb]{\smash{3}}}%
    \put(0.99452469,0.08404003){\color[rgb]{0,0,0}\makebox(0,0)[lb]{\smash{4}}}%
    \put(0.29941861,0.083132){\color[rgb]{0,0,0}\makebox(0,0)[lb]{\smash{}}}%
    \put(0.31425287,0.07643827){\color[rgb]{0,0,0}\makebox(0,0)[lb]{\smash{$\underset{\alpha'\to0}{\longrightarrow}$}}}%
    \put(0.68448513,0.07255017){\color[rgb]{0,0,0}\makebox(0,0)[lb]{\smash{$\equiv$}}}%
  \end{picture}%
\endgroup%

%% file: 3loop-example.pdf_tex
\begingroup%
  \makeatletter%
  \providecommand\color[2][]{%
    \errmessage{(Inkscape) Color is used for the text in Inkscape, but the package 'color.sty' is not loaded}%
    \renewcommand\color[2][]{}%
  }%
  \providecommand\transparent[1]{%
    \errmessage{(Inkscape) Transparency is used (non-zero) for the text in Inkscape, but the package 'transparent.sty' is not loaded}%
    \renewcommand\transparent[1]{}%
  }%
  \providecommand\rotatebox[2]{#2}%
  \ifx\svgwidth\undefined%
    \setlength{\unitlength}{375.97007125bp}%
    \ifx\svgscale\undefined%
      \relax%
    \else%
      \setlength{\unitlength}{\unitlength * \real{\svgscale}}%
    \fi%
  \else%
    \setlength{\unitlength}{\svgwidth}%
  \fi%
  \global\let\svgwidth\undefined%
  \global\let\svgscale\undefined%
  \makeatother%
  \begin{picture}(1,0.20793744)%
    \put(0,0){\includegraphics[width=\unitlength,page=1]{3loop-example.pdf}}%
    \put(0.60388351,0.05587134){\color[rgb]{0,0,0}\makebox(0,0)[lb]{\smash{$\ell_1$}}}%
    \put(0.69812175,0.06282319){\color[rgb]{0,0,0}\makebox(0,0)[lb]{\smash{$\ell_3$}}}%
    \put(0,0){\includegraphics[width=\unitlength,page=2]{3loop-example.pdf}}%
    \put(0.54189344,0.19142319){\color[rgb]{0,0,0}\makebox(0,0)[lb]{\smash{$\ell_2 +k_1 +k_2$}}}%
    \put(0.84478901,0.06130736){\color[rgb]{0,0,0}\makebox(0,0)[lb]{\smash{$\ell_1$}}}%
    \put(0,0){\includegraphics[width=\unitlength,page=3]{3loop-example.pdf}}%
    \put(0.16325238,0.00020606){\color[rgb]{0,0,0}\makebox(0,0)[lb]{\smash{P}}}%
    \put(0.65186756,0.04549586){\color[rgb]{0,0,0}\makebox(0,0)[lb]{\smash{P}}}%
    \put(0.87718299,0.05648064){\color[rgb]{0,0,0}\makebox(0,0)[lb]{\smash{P}}}%
    \put(0,0){\includegraphics[width=\unitlength,page=4]{3loop-example.pdf}}%
    \put(0.09688069,0.05764259){\color[rgb]{0,0,0}\makebox(0,0)[lb]{\smash{$\color{blue}a_1$}}}%
    \put(0.14698752,0.07407079){\color[rgb]{0,0,0}\makebox(0,0)[lb]{\smash{$\color{blue} a_2$}}}%
    \put(0.23110754,0.05553411){\color[rgb]{0,0,0}\makebox(0,0)[lb]{\smash{$\color{blue}a_3$}}}%
    \put(0.02757034,0.02047312){\color[rgb]{0,0,0}\makebox(0,0)[lb]{\smash{1}}}%
    \put(0.00905495,0.18646426){\color[rgb]{0,0,0}\makebox(0,0)[lb]{\smash{2}}}%
    \put(0.34418062,0.18899033){\color[rgb]{0,0,0}\makebox(0,0)[lb]{\smash{3}}}%
    \put(0.37828261,0.05805554){\color[rgb]{0,0,0}\makebox(0,0)[lb]{\smash{4}}}%
    \put(0,0){\includegraphics[width=\unitlength,page=5]{3loop-example.pdf}}%
    \put(0.58073197,0.07477826){\color[rgb]{0,0,0}\makebox(0,0)[lb]{\smash{1}}}%
    \put(0.56471992,0.12335831){\color[rgb]{0,0,0}\makebox(0,0)[lb]{\smash{2}}}%
    \put(0.72116837,0.17900228){\color[rgb]{0,0,0}\makebox(0,0)[lb]{\smash{3}}}%
    \put(0.73820242,0.12505634){\color[rgb]{0,0,0}\makebox(0,0)[lb]{\smash{4}}}%
    \put(0.80973175,0.06264861){\color[rgb]{0,0,0}\makebox(0,0)[lb]{\smash{1}}}%
    \put(0.80498514,0.18406315){\color[rgb]{0,0,0}\makebox(0,0)[lb]{\smash{2}}}%
    \put(0.98972474,0.18252909){\color[rgb]{0,0,0}\makebox(0,0)[lb]{\smash{3}}}%
    \put(0.99463288,0.06344126){\color[rgb]{0,0,0}\makebox(0,0)[lb]{\smash{4}}}%
    \put(0.90657086,0.06154717){\color[rgb]{0,0,0}\makebox(0,0)[lb]{\smash{$\ell_2$}}}%
    \put(0.94924348,0.06026419){\color[rgb]{0,0,0}\makebox(0,0)[lb]{\smash{$\ell_3$}}}%
    \put(0,0){\includegraphics[width=\unitlength,page=6]{3loop-example.pdf}}%
  \end{picture}%
\endgroup%

%% file: loop-momentum.bbl
\providecommand{\href}[2]{#2}\begingroup\raggedright\begin{thebibliography}{10}

\bibitem{Shapiro:1972ph}
J.~A. Shapiro, {\it {Loop Graph in the Dual Tube Model}},  {\em Phys. Rev.}
  {\bf D5} (1972) 1945--1948.

\bibitem{Verlinde:1986kw}
E.~P. Verlinde and H.~L. Verlinde, {\it {Chiral Bosonization, Determinants and
  the String Partition Function}},  {\em Nucl. Phys.} {\bf B288} (1987) 357.
  [,244(1986)].

\bibitem{Verlinde:1987sd}
E.~P. Verlinde and H.~L. Verlinde, {\it {Multiloop Calculations in Covariant
  Superstring Theory}},  {\em Phys.Lett.} {\bf B192} (1987) 95.

\bibitem{DHoker:1988pdl}
E.~D'Hoker and D.~H. Phong, {\it {The Geometry of String Perturbation Theory}},
   {\em Rev. Mod. Phys.} {\bf 60} (1988) 917.

\bibitem{DHoker:1989cxq}
E.~D'Hoker and D.~H. Phong, {\it {Conformal Scalar Fields and Chiral Splitting
  on Superriemann Surfaces}},  {\em Commun. Math. Phys.} {\bf 125} (1989) 469.

\bibitem{Tourkine:2016bak}
P.~Tourkine and P.~Vanhove, {\it {Higher-Loop Amplitude Monodromy Relations in
  String and Gauge Theory}},  {\em Phys. Rev. Lett.} {\bf 117} (2016), no.~21
  211601, [\href{http://xxx.lanl.gov/abs/1608.0166}{{\tt arXiv:1608.0166}}].

\bibitem{Plahte:1970wy}
E.~Plahte, {\it {Symmetry Properties of Dual Tree-Graph N-Point Amplitudes}},
  {\em Nuovo Cim.} {\bf A66} (1970) 713--733.

\bibitem{BjerrumBohr:2009rd}
N.~E.~J. Bjerrum-Bohr, P.~H. Damgaard, and P.~Vanhove, {\it {Minimal Basis for
  Gauge Theory Amplitudes}},  {\em Phys. Rev. Lett.} {\bf 103} (2009) 161602,
  [\href{http://xxx.lanl.gov/abs/0907.1425}{{\tt arXiv:0907.1425}}].

\bibitem{Stieberger:2009hq}
S.~Stieberger, {\it {Open \&amp; Closed vs. Pure Open String Disk Amplitudes}},
   \href{http://xxx.lanl.gov/abs/0907.2211}{{\tt arXiv:0907.2211}}.

\bibitem{Bern:2008qj}
Z.~Bern, J.~J.~M. Carrasco, and H.~Johansson, {\it {New Relations for
  Gauge-Theory Amplitudes}},  {\em Phys. Rev.} {\bf D78} (2008) 085011,
  [\href{http://xxx.lanl.gov/abs/0805.3993}{{\tt arXiv:0805.3993}}].

\bibitem{Bern:2010ue}
Z.~Bern, J.~J.~M. Carrasco, and H.~Johansson, {\it {Perturbative Quantum
  Gravity as a Double Copy of Gauge Theory}},  {\em Phys. Rev. Lett.} {\bf 105}
  (2010) 061602, [\href{http://xxx.lanl.gov/abs/1004.0476}{{\tt
  arXiv:1004.0476}}].

\bibitem{Bern:2018jmv}
Z.~Bern, J.~J. Carrasco, W.-M. Chen, A.~Edison, H.~Johansson,
  J.~Parra-Martinez, R.~Roiban, and M.~Zeng, {\it {Ultraviolet Properties of
  $\mathcal N = 8$ Supergravity at Five Loops}},  {\em Phys. Rev.} {\bf D98}
  (2018), no.~8 086021, [\href{http://xxx.lanl.gov/abs/1804.0931}{{\tt
  arXiv:1804.0931}}].

\bibitem{Feng:2011fja}
B.~Feng, Y.~Jia, and R.~Huang, {\it {Relations of Loop Partial Amplitudes in
  Gauge Theory by Unitarity Cut Method}},  {\em Nucl. Phys.} {\bf B854} (2012)
  243--275, [\href{http://xxx.lanl.gov/abs/1105.0334}{{\tt arXiv:1105.0334}}].

\bibitem{Feng:2010my}
B.~Feng, R.~Huang, and Y.~Jia, {\it {Gauge Amplitude Identities by On-Shell
  Recursion Relation in S-Matrix Program}},  {\em Phys. Lett.} {\bf B695}
  (2011) 350--353, [\href{http://xxx.lanl.gov/abs/1004.3417}{{\tt
  arXiv:1004.3417}}].

\bibitem{Boels:2011tp}
R.~H. Boels and R.~S. Isermann, {\it {New Relations for Scattering Amplitudes
  in Yang-Mills Theory at Loop Level}},  {\em Phys. Rev.} {\bf D85} (2012)
  021701, [\href{http://xxx.lanl.gov/abs/1109.5888}{{\tt arXiv:1109.5888}}].

\bibitem{Boels:2011mn}
R.~H. Boels and R.~S. Isermann, {\it {Yang-Mills Amplitude Relations at Loop
  Level from Non-Adjacent Bcfw Shifts}},  {\em JHEP} {\bf 03} (2012) 051,
  [\href{http://xxx.lanl.gov/abs/1110.4462}{{\tt arXiv:1110.4462}}].

\bibitem{Du:2012mt}
Y.-J. Du and H.~Luo, {\it {On General Bcj Relation at One-Loop Level in
  Yang-Mills Theory}},  {\em JHEP} {\bf 01} (2013) 129,
  [\href{http://xxx.lanl.gov/abs/1207.4549}{{\tt arXiv:1207.4549}}].

\bibitem{Chiodaroli:2017ngp}
M.~Chiodaroli, M.~Gunaydin, H.~Johansson, and R.~Roiban, {\it {Explicit
  Formulae for Yang-Mills-Einstein Amplitudes from the Double Copy}},  {\em
  JHEP} {\bf 07} (2017) 002, [\href{http://xxx.lanl.gov/abs/1703.0042}{{\tt
  arXiv:1703.0042}}].

\bibitem{Green:2012pqa}
M.~B. Green, J.~H. Schwarz, and E.~Witten, {\em {Superstring Theory Vol. 2}}.
\newblock Cambridge Monographs on Mathematical Physics. Cambridge University
  Press, 2012.

\bibitem{Skliros:2016fqs}
D.~P. Skliros, E.~J. Copeland, and P.~M. Saffin, {\it {Highly Excited Strings
  I: Generating Function}},  {\em Nucl. Phys.} {\bf B916} (2017) 143--207,
  [\href{http://xxx.lanl.gov/abs/1611.0649}{{\tt arXiv:1611.0649}}].

\bibitem{farkas}
H.~M. Farkas and I.~Kra, {\em Riemann surfaces}, vol.~71.
\newblock Springer-Verlag, 1980.

\bibitem{Casali:2014hfa}
E.~Casali and P.~Tourkine, {\it {Infrared behaviour of the one-loop scattering
  equations and supergravity integrands}},  {\em JHEP} {\bf 04} (2015) 013,
  [\href{http://xxx.lanl.gov/abs/1412.3787}{{\tt arXiv:1412.3787}}].

\bibitem{Bianchi:1989du}
M.~Bianchi and A.~Sagnotti, {\it {Open Strings and the Relative Modular
  Group}},  {\em Phys. Lett.} {\bf B231} (1989) 389--396.

\bibitem{Scherk:1971xy}
J.~Scherk, {\it {Zero-Slope Limit of the Dual Resonance Model}},  {\em Nucl.
  Phys.} {\bf B31} (1971) 222--234.

\bibitem{Tourkine:2013rda}
P.~Tourkine, {\it {Tropical Amplitudes}},  {\em Annales Henri Poincare} {\bf
  18} (2017), no.~6 2199--2249, [\href{http://xxx.lanl.gov/abs/1309.3551}{{\tt
  arXiv:1309.3551}}].

\bibitem{Fay}
J.~D. Fay, {\em Theta functions on Riemann surfaces}, vol.~352.
\newblock Springer, 1973.

\bibitem{DHoker:2017pvk}
E.~D'Hoker, M.~B. Green, and B.~Pioline, {\it {Higher Genus Modular Graph
  Functions, String Invariants, and Their Exact Asymptotics}},
  \href{http://xxx.lanl.gov/abs/1712.0613}{{\tt arXiv:1712.0613}}.

\bibitem{Dai:2006vj}
P.~Dai and W.~Siegel, {\it {Worldline Green Functions for Arbitrary Feynman
  Diagrams}},  {\em Nucl. Phys.} {\bf B770} (2007) 107--122,
  [\href{http://xxx.lanl.gov/abs/hep-th/0608062}{{\tt hep-th/0608062}}].

\bibitem{Hohenegger:2017kqy}
S.~Hohenegger and S.~Stieberger, {\it {Monodromy Relations in Higher-Loop
  String Amplitudes}},  \href{http://xxx.lanl.gov/abs/1702.0496}{{\tt
  arXiv:1702.0496}}.

\bibitem{Ochirov:2017jby}
A.~Ochirov, P.~Tourkine, and P.~Vanhove, {\it {One-Loop Monodromy Relations on
  Single Cuts}},  {\em JHEP} {\bf 10} (2017) 105,
  [\href{http://xxx.lanl.gov/abs/1707.0577}{{\tt arXiv:1707.0577}}].

\bibitem{D'Hoker:2002gw}
E.~D'Hoker and D.~H. Phong, {\it {Lectures on Two Loop Superstrings}},  {\em
  Conf. Proc.} {\bf C0208124} (2002) 85--123,
  [\href{http://xxx.lanl.gov/abs/hep-th/0211111}{{\tt hep-th/0211111}}].
  [,85(2002)].

\bibitem{Green:2008bf}
M.~B. Green, J.~G. Russo, and P.~Vanhove, {\it {Modular Properties of Two-Loop
  Maximal Supergravity and Connections with String Theory}},  {\em JHEP} {\bf
  07} (2008) 126, [\href{http://xxx.lanl.gov/abs/0807.0389}{{\tt
  arXiv:0807.0389}}].

\bibitem{Bern:1997nh}
Z.~Bern, J.~S. Rozowsky, and B.~Yan, {\it {Two Loop Four Gluon Amplitudes in
  ${\mathcal{N}}\!=4$ Superyang-Mills}},  {\em Phys. Lett.} {\bf B401} (1997)
  273--282, [\href{http://xxx.lanl.gov/abs/hep-ph/9702424}{{\tt
  hep-ph/9702424}}].

\bibitem{Bern:2017ucb}
Z.~Bern, J.~J.~M. Carrasco, W.-M. Chen, H.~Johansson, R.~Roiban, and M.~Zeng,
  {\it {Five-loop four-point integrand of $N=8$ supergravity as a generalized
  double copy}},  {\em Phys. Rev.} {\bf D96} (2017), no.~12 126012,
  [\href{http://xxx.lanl.gov/abs/1708.0680}{{\tt arXiv:1708.0680}}].

\bibitem{ArkaniHamed:2010kv}
N.~Arkani-Hamed, J.~L. Bourjaily, F.~Cachazo, S.~Caron-Huot, and J.~Trnka, {\it
  {The All-Loop Integrand for Scattering Amplitudes in Planar
  ${\mathcal{N}}\!=4$ Sym}},  {\em JHEP} {\bf 01} (2011) 041,
  [\href{http://xxx.lanl.gov/abs/1008.2958}{{\tt arXiv:1008.2958}}].

\bibitem{Arkani-Hamed:2013jha}
N.~Arkani-Hamed and J.~Trnka, {\it {The Amplituhedron}},  {\em JHEP} {\bf 10}
  (2014) 030, [\href{http://xxx.lanl.gov/abs/1312.2007}{{\tt
  arXiv:1312.2007}}].

\bibitem{Herrmann:2016qea}
E.~Herrmann and J.~Trnka, {\it {Gravity On-Shell Diagrams}},  {\em JHEP} {\bf
  11} (2016) 136, [\href{http://xxx.lanl.gov/abs/1604.0347}{{\tt
  arXiv:1604.0347}}].

\bibitem{Mason:2013sva}
L.~Mason and D.~Skinner, {\it {Ambitwistor strings and the scattering
  equations}},  {\em JHEP} {\bf 07} (2014) 048,
  [\href{http://xxx.lanl.gov/abs/1311.2564}{{\tt arXiv:1311.2564}}].

\bibitem{Cachazo:2013iea}
F.~Cachazo, S.~He, and E.~Y. Yuan, {\it {Scattering of Massless Particles:
  Scalars, Gluons and Gravitons}},  {\em JHEP} {\bf 1407} (2014) 033,
  [\href{http://xxx.lanl.gov/abs/1309.0885}{{\tt arXiv:1309.0885}}].

\bibitem{Adamo:2013tsa}
T.~Adamo, E.~Casali, and D.~Skinner, {\it {Ambitwistor strings and the
  scattering equations at one loop}},  {\em JHEP} {\bf 04} (2014) 104,
  [\href{http://xxx.lanl.gov/abs/1312.3828}{{\tt arXiv:1312.3828}}].

\bibitem{Ohmori:2015sha}
K.~Ohmori, {\it {Worldsheet Geometries of Ambitwistor String}},  {\em JHEP}
  {\bf 06} (2015) 075, [\href{http://xxx.lanl.gov/abs/1504.0267}{{\tt
  arXiv:1504.0267}}].

\bibitem{Geyer:2015bja}
Y.~Geyer, L.~Mason, R.~Monteiro, and P.~Tourkine, {\it {Loop Integrands for
  Scattering Amplitudes from the Riemann Sphere}},  {\em Phys. Rev. Lett.} {\bf
  115} (2015), no.~12 121603, [\href{http://xxx.lanl.gov/abs/1507.0032}{{\tt
  arXiv:1507.0032}}].

\bibitem{Geyer:2015jch}
Y.~Geyer, L.~Mason, R.~Monteiro, and P.~Tourkine, {\it {One-loop amplitudes on
  the Riemann sphere}},  {\em JHEP} {\bf 03} (2016) 114,
  [\href{http://xxx.lanl.gov/abs/1511.0631}{{\tt arXiv:1511.0631}}].

\bibitem{Cachazo:2015aol}
F.~Cachazo, S.~He, and E.~Y. Yuan, {\it {One-Loop Corrections from Higher
  Dimensional Tree Amplitudes}},  \href{http://xxx.lanl.gov/abs/1512.0500}{{\tt
  arXiv:1512.0500}}.

\bibitem{He:2015yua}
S.~He and E.~Y. Yuan, {\it {One-Loop Scattering Equations and Amplitudes from
  Forward Limit}},  {\em Phys. Rev.} {\bf D92} (2015), no.~10 105004,
  [\href{http://xxx.lanl.gov/abs/1508.0602}{{\tt arXiv:1508.0602}}].

\bibitem{He:2015wgf}
S.~He, R.~Monteiro, and O.~SCHLotterer, {\it {String-Inspired Bcj Numerators
  for One-Loop Mhv Amplitudes}},  {\em JHEP} {\bf 01} (2016) 171,
  [\href{http://xxx.lanl.gov/abs/1507.0628}{{\tt arXiv:1507.0628}}].

\bibitem{Geyer:2016wjx}
Y.~Geyer, L.~Mason, R.~Monteiro, and P.~Tourkine, {\it {Two-Loop Scattering
  Amplitudes from the Riemann Sphere}},  {\em Phys. Rev.} {\bf D94} (2016),
  no.~12 125029, [\href{http://xxx.lanl.gov/abs/1607.0888}{{\tt
  arXiv:1607.0888}}].

\bibitem{Geyer:2018xwu}
Y.~Geyer and R.~Monteiro, {\it {Two-Loop Scattering Amplitudes from Ambitwistor
  Strings: from Genus Two to the Nodal Riemann Sphere}},  {\em JHEP} {\bf 11}
  (2018) 008, [\href{http://xxx.lanl.gov/abs/1805.0534}{{\tt
  arXiv:1805.0534}}].

\bibitem{Kawai:1985xq}
H.~Kawai, D.~C. Lewellen, and S.~H.~H. Tye, {\it {A Relation Between Tree
  Amplitudes of Closed and Open Strings}},  {\em Nucl. Phys.} {\bf B269} (1986)
  1--23.

\bibitem{BjerrumBohr:2010hn}
N.~E.~J. Bjerrum-Bohr, P.~H. Damgaard, T.~Sondergaard, and P.~Vanhove, {\it
  {The Momentum Kernel of Gauge and Gravity Theories}},  {\em JHEP} {\bf 01}
  (2011) 001, [\href{http://xxx.lanl.gov/abs/1010.3933}{{\tt
  arXiv:1010.3933}}].

\bibitem{Mizera:2017cqs}
S.~Mizera, {\it {Combinatorics and Topology of Kawai-Lewellen-Tye Relations}},
  \href{http://xxx.lanl.gov/abs/1706.0852}{{\tt arXiv:1706.0852}}.

\bibitem{Mizera:2017rqa}
S.~Mizera, {\it {Scattering Amplitudes from Intersection Theory}},
  \href{http://xxx.lanl.gov/abs/1711.0046}{{\tt arXiv:1711.0046}}.

\bibitem{Mastrolia:2018uzb}
P.~Mastrolia and S.~Mizera, {\it {Feynman Integrals and Intersection Theory}},
  \href{http://xxx.lanl.gov/abs/1810.0381}{{\tt arXiv:1810.0381}}.

\bibitem{Mafra:2017ioj}
C.~R. Mafra and O.~Schlotterer, {\it {Double-Copy Structure of One-Loop
  Open-String Amplitudes}},  {\em Phys. Rev. Lett.} {\bf 121} (2018), no.~1
  011601, [\href{http://xxx.lanl.gov/abs/1711.0910}{{\tt arXiv:1711.0910}}].

\bibitem{Mafra:2018nla}
C.~R. Mafra and O.~SCHLotterer, {\it {Towards the N-Point One-Loop Superstring
  Amplitude I: Pure Spinors and Superfield Kinematics}},
  \href{http://xxx.lanl.gov/abs/1812.1096}{{\tt arXiv:1812.1096}}.

\bibitem{Mafra:2018qqe}
C.~R. Mafra and O.~SCHLotterer, {\it {Towards the N-Point One-Loop Superstring
  Amplitude Iii: One-Loop Correlators and Their Double-Copy Structure}},
  \href{http://xxx.lanl.gov/abs/1812.1097}{{\tt arXiv:1812.1097}}.

\bibitem{Mafra:2018pll}
C.~R. Mafra and O.~SCHLotterer, {\it {Towards the N-Point One-Loop Superstring
  Amplitude Ii: Worldsheet Functions and Their Duality to Kinematics}},
  \href{http://xxx.lanl.gov/abs/1812.1097}{{\tt arXiv:1812.1097}}.

\bibitem{Siegel:2015axg}
W.~Siegel, {\it {Amplitudes for Left-Handed Strings}},
  \href{http://xxx.lanl.gov/abs/1512.0256}{{\tt arXiv:1512.0256}}.

\bibitem{Huang:2016bdd}
Y.-t. Huang, W.~Siegel, and E.~Y. Yuan, {\it {Factorization of Chiral String
  Amplitudes}},  {\em JHEP} {\bf 09} (2016) 101,
  [\href{http://xxx.lanl.gov/abs/1603.0258}{{\tt arXiv:1603.0258}}].

\bibitem{Leite:2016fno}
M.~M. Leite and W.~Siegel, {\it {Chiral Closed Strings: Four Massless States
  Scattering Amplitude}},  \href{http://xxx.lanl.gov/abs/1610.0205}{{\tt
  arXiv:1610.0205}}.

\bibitem{Casali:2017mss}
E.~Casali and P.~Tourkine, {\it {Winding Modes of Tensionful Ambitwistor
  Strings}},  \href{http://xxx.lanl.gov/abs/1710.0124}{{\tt arXiv:1710.0124}}.

\bibitem{Hwang:1998gs}
S.~Hwang, R.~Marnelius, and P.~Saltsidis, {\it {A General BRST Approach to
  String Theories with Zeta Function Regularizations}},  {\em J. Math. Phys.}
  {\bf 40} (1999) 4639--4657,
  [\href{http://xxx.lanl.gov/abs/hep-th/9804003}{{\tt hep-th/9804003}}].

\bibitem{Gamboa:1989zc}
J.~Gamboa, C.~Ramirez, and M.~Ruiz-Altaba, {\it {Null Spinning Strings}},  {\em
  Nucl. Phys.} {\bf B338} (1990) 143--187.

\bibitem{Gamboa:1989px}
J.~Gamboa, C.~Ramirez, and M.~Ruiz-Altaba, {\it {Quantum Null (Super)Strings}},
   {\em Phys. Lett.} {\bf B225} (1989) 335--339.

\bibitem{Casali:2017zkz}
E.~Casali, Y.~Herfray, and P.~Tourkine, {\it {The Complex Null String, Galilean
  Conformal Algebra and Scattering Equations}},  {\em JHEP} {\bf 10} (2017)
  164, [\href{http://xxx.lanl.gov/abs/1707.0990}{{\tt arXiv:1707.0990}}].

\bibitem{Casali:2016atr}
E.~Casali and P.~Tourkine, {\it {On the Null Origin of the Ambitwistor
  String}},  \href{http://xxx.lanl.gov/abs/1606.0563}{{\tt arXiv:1606.0563}}.

\bibitem{Gross:1987ar}
D.~J. Gross and P.~F. Mende, {\it {String Theory Beyond the Planck Scale}},
  {\em Nucl. Phys.} {\bf B303} (1988) 407--454.

\bibitem{Kiritsis:2007zza}
E.~Kiritsis, {\em {String Theory in a Nutshell}}.
\newblock 2007.

\end{thebibliography}\endgroup
